\documentclass[twocolumn,amsmath,amssymb,pra,longbibliography,superscriptaddress]{revtex4-2}

\usepackage{graphicx}
\usepackage{amsmath,amsfonts,amssymb,amsthm}
\usepackage{bm}
\usepackage{bbm}
\usepackage[colorlinks]{hyperref}
\usepackage{xcolor}
\usepackage{placeins}

\graphicspath{{figures/}}

\begin{document}

\author{Juha Lepp\"akangas}
\affiliation{HQS Quantum Simulations GmbH, Rintheimer Str. 23, 76131 Karlsruhe, Germany}

\author{Michael Marthaler}
\affiliation{HQS Quantum Simulations GmbH, Rintheimer Str. 23, 76131 Karlsruhe, Germany}

\title{Direct-decision quantum sensing for classifying transient NMR signals}

\begin{abstract}
Task-oriented quantum sensing aims to extract decision-relevant information
directly, without first reconstructing the full signal.
We study this approach for mixture-threshold classification of transient
NMR-like signals generated from molecular spin
Hamiltonians, motivated by small-volume molecular sensing.
Using analytical models and numerical simulations, we compare
classical coil readout followed by matched-filter classification
with quantum computational sensing (QCS), in which trainable
controls interleaved with signal-driven sensor-qubit evolution
map the transient directly to a decision score.
We consider both the sensor-ensemble limit and strict
single-sensor single-shot readout.
Increasing QCS control depth can improve robustness and
class-dependent output contrast. Ensemble accuracy can rise
rapidly for some waveform pairs, whereas single-shot
correctness improves more gradually because one projective
outcome must distinguish the classes.
Repeated acquisitions improve both routes by averaging independent
fluctuations, while disorder-aware training mitigates quasistatic
inhomogeneity in the modeled sensor ensemble.
QCS can outperform inductive readout under the modeled
conditions, for example when coil transfer is weak or receiver
noise limits coil performance. However, the simulated QCS
protocols at lower effective sensor coupling require long
sensing intervals, whereas short-time protocols require a
much stronger effective drive at the sensor. Practical
advantage therefore depends on jointly attainable
source-signal strength, signal-to-sensor transfer, qubit
coherence, and readout resources.
\end{abstract}

\maketitle


\section{Introduction}
\label{sec:introduction}

Nuclear magnetic resonance (NMR) spectroscopy is a
central tool for molecular identification, mixture analysis, and materials
characterization. Conventional NMR combines inductive detection, repeated
acquisition, and spectral reconstruction to obtain detailed molecular
information. At small sample volumes, however, inductive signals become weak,
efficient placement of a coil or resonator may be difficult, and inductive
readout does not naturally provide highly localized field sensitivity.
Moreover, the experimental objective is often a specific decision rather
than full spectral reconstruction, for example whether a mixture component
exceeds a prescribed threshold. This motivates task-oriented sensing
strategies that extract only the information needed for the final inference
problem.

Quantum sensors, including nitrogen-vacancy (NV) centers in diamond,
provide local access to NMR-like signals from small sample volumes~\cite{staudacher2013five_nanometer_nmr,Mamin2013,
Schirhagl2014,Degen2017QuantumSensing}.
Synchronized readout and quantum-memory protocols can extend spectral
resolution beyond the bare electronic-spin coherence
time~\cite{Schmitt2017Qdyne,Glenn2018CASR,Aslam2017Memory,
zaiser2016enhancing}. Advances in coherent control and pulse
optimization have also enabled more flexible control-assisted sensing
protocols~\cite{Jauch2025,Poggiali2018OptimalControl,Joas2025}.
This raises the question of whether a quantum sensor can use a
transient NMR signal to make a task-relevant decision directly,
without first reconstructing the complete waveform or spectrum.

This question is part of the broader program of \emph{quantum computational
sensing} (QCS), in which signal-driven evolution is interleaved with
coherent processing so that the final measurement encodes a desired function
of the signal rather than an intermediate signal estimate
~\cite{Khan2025QCSAdvantage,Khan2026QCS,allen2025quantum_computing_enhanced_sensing}.
There is a close analogy to quantum-signal
processing~\cite{Low2016CompositeGates,Low2017QSP,
Haah2019QSPDecomposition,Dong2021QSPPhases}, which
instead considers repeated interleaving operations on a
time-independent signal.
Recent QCS work has demonstrated displacement classification in an
oscillator-qubit system~\cite{Prabhu2026QCDS}, studied
single-qubit-assisted learning of signal features~\cite{Kannan2026QFS},
and realized magnetic-field classification with a superconducting qubit~\cite{Sen2026QCS}.
Such studies have compared QCS with task-appropriate
conventional sensing baselines, including Ramsey sensing
and dynamical decoupling followed by classical processing~\cite{Sen2026QCS}.

Here we study QCS for structured transients generated from
molecular NMR spin Hamiltonians, focusing on mixture-threshold
classification from the time-domain waveform. To assess direct
quantum processing against conventional inductive detection
under specified detector models, we compare QCS with noisy coil
readout followed by classical waveform classification.

In the coil route, the source signal is transferred to an
inductive receiver and classified from the recorded complex
waveform, which also contains receiver noise.
In QCS, a signal drawn from the same noisy source model
drives a sensor qubit during a sequence of trainable controls.
We distinguish two QCS readout limits. In expectation-value QCS,
formally corresponding to an ideal ensemble of identically
controlled sensors ($N_s\to\infty$), projection noise vanishes:
the final expectation is available exactly for each source-noise
realization. We also call this the \emph{deterministic} or
\emph{ensemble} limit. At $N_s=1$, a single acquisition instead
provides one projective outcome.
Independent acquisitions ($N_r>1$) average source-noise
fluctuations in both routes. They also average receiver noise
in the coil route and projective-readout noise in finite-sensor
QCS.

A separate one-vs-rest molecular-library task is presented in
the Appendix. NMR provides a practically motivated source of
structured transients, but the framework also applies to
non-NMR transient classification tasks under the stated signal,
coupling, and noise assumptions. Our analysis gives an explicit
accuracy relation for the coil baseline, bounds attainable QCS
contrast, and describes the effects of source noise, repeated
acquisitions, and finite-sensor readout.

Across the tested molecular mixture pairs, QCS can achieve
well-above-chance accuracy, although its coupling dependence is
pair dependent. Exact expectation-value classification and
single-sensor, single-shot readout have different requirements:
a small but separable expectation-value response need not yield
informative projective outcomes. Repetition improves both routes
when independent transients are available.
Under the specified conditions, QCS can outperform inductive
readout when coil transfer is weak or receiver noise is limiting.
Practical advantage remains conditional, however: the simulated
protocols with weaker effective drives at the sensor require
demanding uninterrupted sensing intervals, whereas short-time
protocols require much stronger drives. Whether either regime
can be realized for a molecular sample depends on the
signal-to-sensor interface and the sensor's coherence under
the required controls.

The paper is structured as follows.
Sec.~\ref{sec:model-and-routes} defines the signal and detector
models, Sec.~\ref{sec:analytics-main} presents the analytical
results, and Sec.~\ref{sec:results} reports the numerical
comparisons. Section~\ref{sec:conclusion} gives the conclusions;
the Appendices provide derivations, simulation details, and
additional benchmarks.

\begin{figure*}[t]
\centering
\includegraphics[width=1.99\columnwidth]{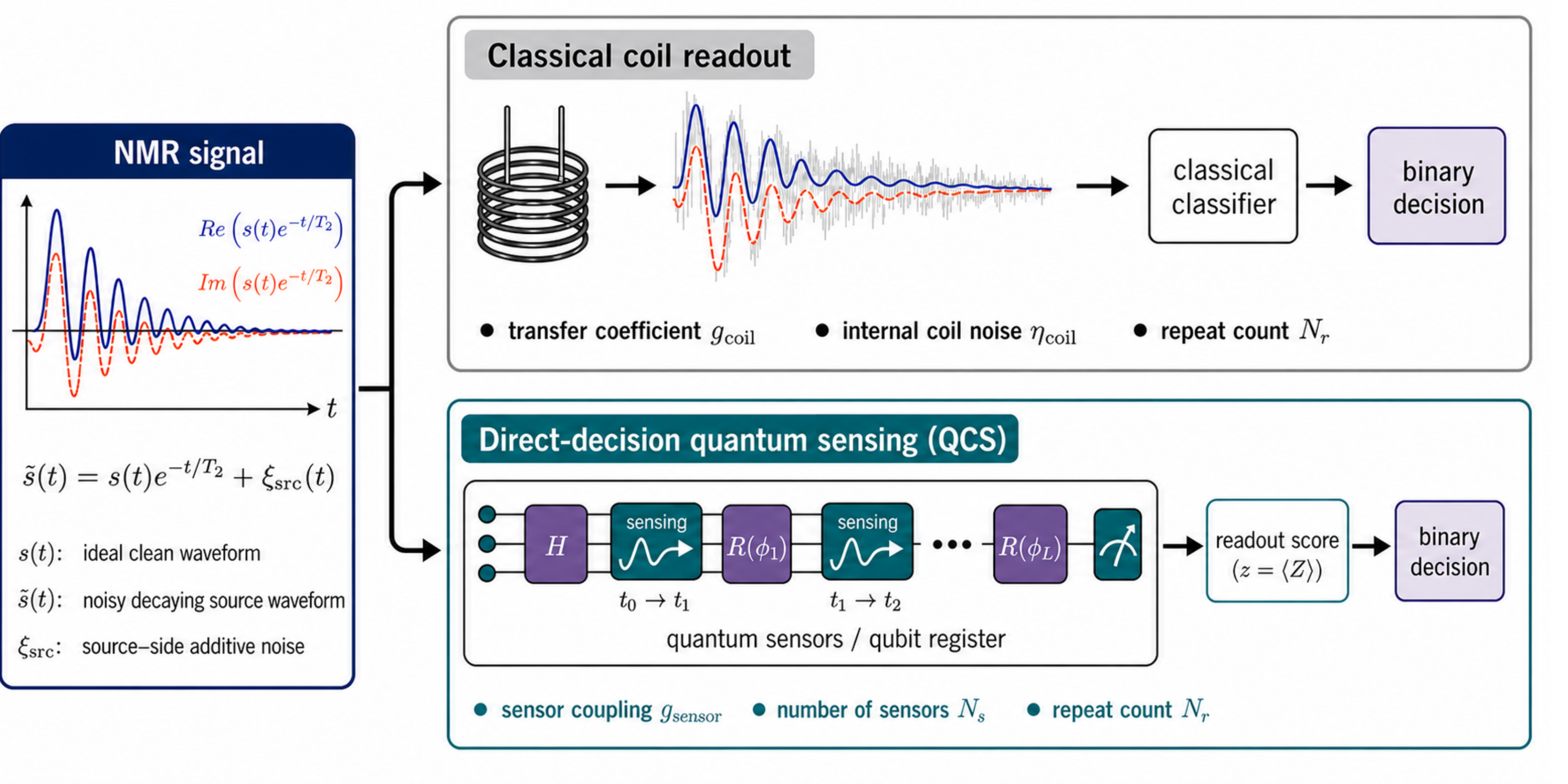}
\caption{
\textbf{Two routes for classifying transient NMR signals.}
Both routes receive the modeled source signal $\tilde s(t)$,
comprising a normalized molecular transient, a $T_2$ decay
envelope, and additive source noise.
\textbf{Top:} coil readout transfers the signal with coefficient
$g_{\rm coil}$, adds receiver noise, and classifies the recorded
complex waveform.
\textbf{Bottom:} quantum computational sensing (QCS) couples
the signal to sensor qubits with strength $g_{\rm sensor}$ and
interleaves sensing with trainable controls. Projective
$Z$-basis readout supplies the final decision score.
Here $N_s$ denotes sensors per acquisition and $N_r$ denotes
independent acquisitions. The limit $N_s\to\infty$ gives an
exact expectation value for each noisy source waveform;
$N_s=1$ gives one projective outcome per acquisition.
Strict single-sensor, single-shot classification has
$N_s=N_r=1$. For $N_r>1$, readouts from independent
acquisitions are combined before classification.
The depicted $H$ gate illustrates equatorial-state
preparation; the mixture simulations instead prepare
$|-i\rangle=(|0\rangle-i|1\rangle)/\sqrt{2}$.
The schematic artwork was generated with assistance from
ChatGPT image-generation tool.
}
  \label{fig:intro_routes}
\end{figure*}


\section{Signal model and compared sensing routes}
\label{sec:model-and-routes}

Fig.~\ref{fig:intro_routes} distinguishes the clean molecular
transient, source fluctuations, and the two detector-specific
processing routes. Both routes receive the same statistical source
signal, but produce different detector outputs.

\subsection{Transient NMR signal and classification task}
\label{sec:signal-model}

Molecule \(m\) generates a complex transient \(s^{(m)}(t)\) from
evolution under a molecular NMR spin
Hamiltonian~\cite{Fratus2025NMRQuantumSimulation, hqs_nmr_parameters}.
Retaining spin-\(\tfrac12\) nuclei, we write
\begin{equation}
H_{\mathrm{mol}}
=
\frac12\sum_i\omega_i Z_i
+
\frac14\sum_{i<j}J_{ij}
\left(X_iX_j+Y_iY_j+Z_iZ_j\right),
\label{eq:molecular-hamiltonian}
\end{equation}
where the resonance offsets \(\omega_i\), defined relative to a
proton rotating frame, and the scalar couplings \(J_{ij}\) are in
angular-frequency units. Equivalently,
\(H_{\mathrm{mol}}=\sum_i\omega_i I_{iz}
+\sum_{i<j}J_{ij}\mathbf I_i\cdot\mathbf I_j\),
with \(I_{i\alpha}=\sigma_{i\alpha}/2\).

For each molecular pair, the transients are represented as complex
envelopes in a common, pair-centered proton rotating frame. At
\(B_0=11.74~\mathrm{T}\), the frame removes a carrier near the
\(^{1}\mathrm H\) Larmor frequency of \(500~\mathrm{MHz}\) from
the waveform representation.

Suppressing the molecular index, we define the noisy source signal
presented to either detector as
\begin{equation}
\tilde s(t)
=
s(t)e^{-t/T_2}+\xi_{\mathrm{src}}(t),
\label{eq:latent-source-waveform}
\end{equation}
where \(T_2\) is a phenomenological source-signal decay time and
\(\xi_{\mathrm{src}}\) is additive source noise. We model its
I- and Q-quadratures (real and imaginary parts) as independent, zero-mean, temporally
white Gaussian fluctuations, with formal continuous-time covariance
\begin{equation}
\begin{aligned}
\mathbb E[\xi_a(t)\xi_b(t')]
&=S_{\rm src}\,\delta_{ab}\delta(t-t'),\\
S_{\rm src}&=\sigma_{\rm src}^{\,2}dt,
\qquad a,b\in\{I,Q\}.
\end{aligned}
\label{eq:source-noise-iq}
\end{equation}
In the simulations, independent samples of each quadrature are drawn
with standard deviation \(\sigma_{\rm src}\) at each time step \(dt\).
Independent acquisitions receive independent source-noise draws.

The principal task is binary mixture-threshold detection. For
normalized molecular transients \(s^{(A)}(t)\) and \(s^{(B)}(t)\)
(Appendix~\ref{app:simulation-protocol}), define
\begin{equation}
s^{(\mathrm{mix})}(t)
=
\alpha s^{(A)}(t)
+
(1-\alpha)s^{(B)}(t).
\label{eq:smix}
\end{equation}
The common decay envelope and source fluctuations are applied
after mixing. The classification label is
\begin{equation}
y=
\begin{cases}
+1,&\alpha\geq\alpha_0,\\
-1,&\alpha<\alpha_0.
\end{cases}
\label{eq:mixture-label}
\end{equation}
Appendix~\ref{app:real_library_identification} considers a separate
one-vs-rest molecular-library task.

\subsection{Classical coil readout}
\label{sec:coil-model}

The coil route records the noisy NMR signal and classifies it
classically. Its effective observation model is
\begin{equation}
c(t)
=
g_{\rm coil}\tilde s(t)+\xi_{\rm coil}(t).
\label{eq:coil-readout}
\end{equation}
Here $g_{\rm coil}$ is an end-to-end signal-transfer
coefficient, and $\xi_{\rm coil}$ is receiver noise independent
of the source noise, with per-quadrature standard deviation
$\eta_{\rm coil}$ on the sampling grid.
The recorded complex trace is passed to a linear
matched-filter-type classifier; its score and threshold
calibration are described in Appendix~\ref{app:coil-analytic-derivation}.
The coefficient $g_{\rm coil}$ incorporates pickup efficiency,
receiver response, and amplification. We treat it as an
effective parameter rather than calculate it for a specified
coil geometry and receiver chain.

\subsection{Quantum computational sensing}
\label{sec:qcs-model}

In QCS~\cite{Khan2025QCSAdvantage,Khan2026QCS}, the
transient signal drives a sensor qubit during a sequence of trainable
controls rather than being recorded first as a classical trace.
We model the effective signal at the sensor as
\begin{equation}
q(t)=g_{\rm sensor}\tilde s(t),
\label{eq:qcs-drive-discrete}
\end{equation}
with signal Hamiltonian
\begin{equation}
H_{\mathrm{sig}}(t)
=
\frac12\left[
\operatorname{Re}q(t)\,X+
\operatorname{Im}q(t)\,Y
\right].
\label{eq:Hsig-qcs-main}
\end{equation}
This is an effective sensor-rotating-frame interaction: the
complex envelope supplies phase-resolved transverse drive
quadratures, and sensor detuning is omitted. Such an $X$--$Y$
form arises for a resonantly driven two-level sensor in a
rotating frame~\cite{Meinel2021Heterodyne}. For a constant
envelope, $|q|$ is the angular rotation rate in our Hamiltonian
convention.

The model assumes an interface that maps the molecular
envelope to the sensor's transverse drive while preserving
its phase and time dependence. The effective drive strength
is set by $g_{\rm sensor}$. Representing the NMR waveform
in a proton rotating frame does not itself implement this
mapping. Requirements for a physical implementation,
including with an NV sensor, are discussed in
Appendix~\ref{app:parameter_interpretation}.

The sensing sequence starts from initialization of the sensor qubit(s).
In all shown simulations, the initial state was chosen to be
\begin{equation}
\lvert -i\rangle
=
\frac{\lvert0\rangle-i\lvert1\rangle}{\sqrt2}.
\end{equation}
For a controlled protocol, we divide the total sensing interval into \(L\) segments,
\(0=t_0<t_1<\cdots<t_L=T\). Signal evolution during segment \(\ell\)
is
\begin{equation}
U_{\mathrm{sig},\ell}
=
\mathcal T\exp\left[
-i\int_{t_{\ell-1}}^{t_\ell}H_{\mathrm{sig}}(t)\,dt
\right].
\label{eq:Usig-qcs-main}
\end{equation}
The complete controlled evolution is
\begin{equation}
U_{\mathrm{QCS}}
=
V_L U_{\mathrm{sig},L}\cdots V_1 U_{\mathrm{sig},1}.
\label{eq:Uqcs-main}
\end{equation}
We parameterize each control as
\begin{equation}
V_\ell(\phi_\ell,\theta_\ell)
=
R_z(\phi_\ell)R_x(\theta_\ell).
\label{eq:qcs-control-gate}
\end{equation}
Controls are modeled as instantaneous and common to all sensor
copies.
Their optimization is described in Appendix~\ref{app:qcs-optimizer}.

After the final control, a projective \(Z\) measurement yields
\(x\in\{+1,-1\}\). If \(z=\langle Z\rangle\), then
\begin{equation}
\mathbb P(x=+1\mid z)=\frac{1+z}{2},
\qquad
\mathbb P(x=-1\mid z)=\frac{1-z}{2}.
\label{eq:qcs-bernoulli}
\end{equation}

For \(N_s\) parallel sensors per acquisition and \(N_r\)
independent acquisitions, the measured average is
\begin{equation}
\bar M
=
\frac{1}{N_rN_s}
\sum_{r=1}^{N_r}\sum_{a=1}^{N_s}x_{r,a}.
\label{eq:qcs-averaged-readout}
\end{equation}
A calibrated decision rule classifies \(\bar M\). For finite
\(N_sN_r\), thresholding an affine function of \(\bar M\) is
equivalent to thresholding the integer count of \(+1\)
outcomes.
The limit
\(N_s\to\infty\) gives the exact expectation for
\emph{each} acquisition, but does not remove source
fluctuations between acquisitions. At \(N_s=N_r=1\), the
decision uses one projective outcome, whose probabilities
are given by Eq.~\eqref{eq:qcs-bernoulli}.


\section{Analytical results}
\label{sec:analytics-main}

Both sensing routes ultimately produce a decision score. Their
classification performance depends on class-dependent score
separation relative to score fluctuations. For the coil route,
this relation follows explicitly from the linear Gaussian-noise model.
For QCS, the score is generated by trained, time-ordered dynamics;
we instead identify a bound on attainable contrast and characterize
the resulting score statistics.

\subsection{Classical coil readout}
\label{sec:analytics-coil}

Let the vector difference between the clean, decayed
waveforms of the molecular pair be
\begin{equation}
\mathbf v
=
\left(
[s_k^{(A)}-s_k^{(B)}]e^{-t_k/T_2}
\right)_{k=0}^{K-1}.
\label{eq:coil-mixture-direction}
\end{equation}
For $\delta=\alpha-\alpha_0$, the deterministic coil waveform differs
from that at the mixture boundary by $g_{\rm coil}\delta\mathbf v$.
The per-quadrature standard deviation of the sampled coil noise is
\begin{equation}
\sigma_{\rm eff}
=
\sqrt{\eta_{\rm coil}^{\,2}
      +g_{\rm coil}^{\,2}\sigma_{\rm src}^{\,2}}.
\label{eq:coil-sigmaeff-main}
\end{equation}
We construct a task-specific, matched-filter-type coil baseline by
projecting the centered observation onto $\mathbf v$
(Appendix~\ref{app:coil-analytic-derivation}). Its ideal
single-acquisition correctness, with the threshold at the physical
mixture boundary, is
\begin{equation}
P_{\rm corr}^{\rm coil}(\delta)
=
\Phi\!\left(|\delta|\mathcal D_{\rm coil}\right),
\qquad
\mathcal D_{\rm coil}
=
\frac{g_{\rm coil}\|\mathbf v\|}{\sigma_{\rm eff}},
\label{eq:coil-local-delta-main}
\end{equation}
where $\Phi$ is the standard normal cumulative distribution
function. 
The numerical classifier uses the same linear-score
direction but calibrates its threshold on separate data.
For the benchmark shown here, this calibration changes
held-out accuracy only slightly relative to the ideal
boundary threshold.

The difference between the pure-component waveforms therefore
does not by itself specify task accuracy: mixtures close to
$\alpha_0$ remain difficult.
Our benchmark samples $\alpha$ continuously on both sides of the
threshold, so aggregate accuracy also depends on the distribution
of $|\delta|$. At weak transfer, receiver noise limits the coil;
at strong transfer, propagated source noise becomes limiting.
Averaging $N_r$ independent acquisitions increases the ideal
boundary-relative discriminability by $\sqrt{N_r}$.

\subsection{Quantum computational sensing}
\label{sec:analytics-qcs}

\subsubsection{Attainable contrast and score separation}

For one source-noise realization $\xi$, let
$z_\xi(\alpha)=\langle Z\rangle$ be the final \emph{physical}
sensor expectation. Exact expectation-value readout makes this
score deterministic given $\xi$, but different source waveforms
produce different scores at the same mixture fraction.

Define the decayed waveform-difference scale
\begin{equation}
D_{\rm sep}(T)
=
\int_0^T
|s^{(A)}(t)-s^{(B)}(t)|e^{-t/T_2}\,dt.
\label{eq:qcs-separation-scale}
\end{equation}
For the same controls, initial state, and source-noise realization
at $\alpha_0+\delta$ and $\alpha_0$,
\begin{equation}
|z_\xi(\alpha_0+\delta)-z_\xi(\alpha_0)|
\leq
g_{\rm sensor}|\delta|D_{\rm sep}(T).
\label{eq:qcs-contrast-bound}
\end{equation}
The bound follows from the difference between the two signal
Hamiltonians and $\|Z\|_{\rm op}=1$.
We call $g_{\rm sensor}|\delta|D_{\rm sep}(T)$ the
boundary-relative class-dependent action. If it is small,
the physical expectation cannot differ substantially from
that of the boundary mixture, regardless of the controls.
A large value permits, but does not guarantee, useful
one-shot contrast: the trained dynamics determine the
response actually achieved.

To describe the achieved deterministic score, define
\begin{equation}
\mu_\alpha=\mathbb E_\xi[z_\xi(\alpha)],
\quad
v_\alpha=\operatorname{Var}_\xi[z_\xi(\alpha)],
\quad
\beta_{\rm QCS}
=
\left.
\frac{\partial\mu_\alpha}{\partial\alpha}
\right|_{\alpha=\alpha_0} 
\label{eq:qcs-local-score-statistics}
\end{equation}
Near $\alpha_0$, if the source-induced score fluctuations are
approximately Gaussian and the threshold is placed at
$\mu_{\alpha_0}$, then
\begin{equation}
P_{\rm corr}^{\rm det}(\delta)
\approx
\Phi\!\left(
\frac{|\beta_{\rm QCS}\delta|}
{\sqrt{v_{\alpha_0+\delta}}}
\right).
\label{eq:qcs-local-probit_main}
\end{equation}
Unlike $\mathcal D_{\rm coil}$, the slope and variance depend
on the trained dynamics. The bound in
Eq.~\eqref{eq:qcs-contrast-bound} constrains possible
contrast, whereas Eq.~\eqref{eq:qcs-local-probit_main}
describes classification from the contrast and fluctuations
actually produced. For $N_r$ independent acquisitions,
averaging the exact expectation scores preserves their mean
and reduces $v_\alpha$ to $v_\alpha/N_r$. Within the local
Gaussian approximation, the argument of $\Phi$ therefore
grows as $\sqrt{N_r}$. The numerical comparison
in Sec.~\ref{sec:results-regimemap} confirms this dependence.

\subsubsection{Source-noise exposure and sensor coherence}

Increasing $g_{\rm sensor}$ raises the class-dependent action
linearly, but also strengthens the modeled fluctuations driving
the qubit. With the per-quadrature source-noise intensity
$S_{\rm src}=\sigma_{\rm src}^{\,2}dt$, define
\begin{equation}
\Gamma_{\rm src}
=
g_{\rm sensor}^{\,2}S_{\rm src}.
\label{eq:qcs-source-noise-rate}
\end{equation}
In the small-noisy-rotation-per-step approximation, the
source-noise averaged state evolves between controls as
\begin{equation}
\frac{d\bar\rho}{dt}
\approx
-i[H_{\rm clean}(t),\bar\rho]
-\frac{\Gamma_{\rm src}}{8}
\left(
[X,[X,\bar\rho]]+[Y,[Y,\bar\rho]]
\right).
\label{eq:qcs-source-averaged-master}
\end{equation}
Without clean drive or controls, its Bloch components obey
\begin{equation}
\dot{\bar x}=-\frac{\Gamma_{\rm src}}{2}\bar x,
\qquad
\dot{\bar y}=-\frac{\Gamma_{\rm src}}{2}\bar y,
\qquad
\dot{\bar z}=-\Gamma_{\rm src}\bar z.
\label{eq:qcs-source-bloch-rates}
\end{equation}
These rates describe exponential attenuation of components of
the \emph{source-noise-averaged} sensor state, even though each
sensor evolution conditioned on a source waveform is unitary.
Thus, source fluctuations alone can limit the useful
interaction time: extending one acquisition accumulates
waveform difference, but also increases exposure to the
noise-induced channel. The same fluctuations generate the
score variance $v_\alpha$ in
Eq.~\eqref{eq:qcs-local-probit_main}; attenuation of mean
separation and variation of individual scores are two
statistics of one noise process, not separate penalties.

The molecular drive and controls mix Bloch components, so
$\Gamma_{\rm src}^{-1}$ is an exposure scale, not a prediction
of the optimal sensing time or a universal decay time for
classification accuracy.
Independent acquisitions can still improve the precision with
which a nonzero remaining response is resolved.
Intrinsic sensor
decoherence, not simulated here, may impose a further or more
restrictive time constraint.
The derivation and numerical checks of the source-noise-exposure effect
are presented in Appendix~\ref{app:source-noise-channel}.

\subsection{Repeated acquisitions and finite sensor number}
\label{sec:repeated_acquisitions}

Consider the measured average $\bar M$ defined in
Eq.~\eqref{eq:qcs-averaged-readout}. We assume that $N_s$
sensors share one source waveform within each acquisition,
while source noise is independently redrawn across the $N_r$
acquisitions.
The law of total variance gives
\begin{equation}
\operatorname{Var}(\bar M\mid\alpha)
=
\frac{1}{N_r}
\left[
v_\alpha+
\frac{\mathbb E_\xi[1-z_\xi(\alpha)^2]}{N_s}
\right].
\label{eq:Nr-Ns-qcs-main}
\end{equation}
The first term is source-induced variation of the final
expectation; the second is finite-sensor projection noise.
Increasing $N_s$ suppresses only the latter, whereas independent
repetition suppresses both. Thus deterministic QCS
($N_s\to\infty$) can still benefit from repeated acquisitions.

For one sensor per acquisition, let $K_+$ be the number of
$+1$ outcomes across $N_r$ acquisitions. Then
$\bar M=2K_+/N_r-1$: thresholding an affine function of
$\bar M$ is equivalently a threshold on the integer count
$K_+$. As $N_r$ grows, this measured average approaches
$\mu_\alpha=\mathbb E_\xi[z_\xi(\alpha)]$. It thus approaches
the source-noise-averaged expectation, not the exact
expectation of any one noisy acquisition.

At $N_s=N_r=1$, the sensor instead yields just one binary
outcome $x=\pm1$. For a fixed mixture and source-noise
realization, and the outcome assignment $\hat y=x$, the
probability of a correct decision is
\begin{equation}
P(x=y\mid\alpha,\xi)
=
\frac{1+yz_\xi(\alpha)}{2},
\label{eq:pcorr-single-main}
\end{equation}
where $y$ is the true label from
Eq.~\eqref{eq:mixture-label}. Calibration may reverse this
assignment. An exact expectation can be thresholded even
when class-dependent differences are small; one projective
outcome is informative only when its probabilities differ
appreciably between classes. Repeated single-sensor outcomes
provide a finer decision score while averaging both source
variation and projection noise.

The numerical results below quantify how control depth,
coupling, sensing duration, repetition, and sensor disorder
affect classification performance.


\section{Numerical results}
\label{sec:results}

We test how control depth, sensing duration, coupling, repeated
acquisitions, and sensor disorder affect mixture-threshold
classification. Throughout, deterministic QCS accuracy classifies
the exact final expectation for each noisy waveform; one-shot
correctness is the expected success probability of one projective
sensor readout. The main-text figures use one representative
molecular pair. Results for additional pairs and a separate
molecular-library task are given in
Appendices~\ref{app:mixture_pair_robustness}
and~\ref{app:real_library_identification}.

QCS curves show means over independent training seeds, typically
20--30 per operating point unless stated otherwise; error bars
indicate the standard error of the mean (SEM). The curves represent
trained performance, not certified global optima.

\begin{figure*}[!t]
\centering
\includegraphics[width=0.95\linewidth]
{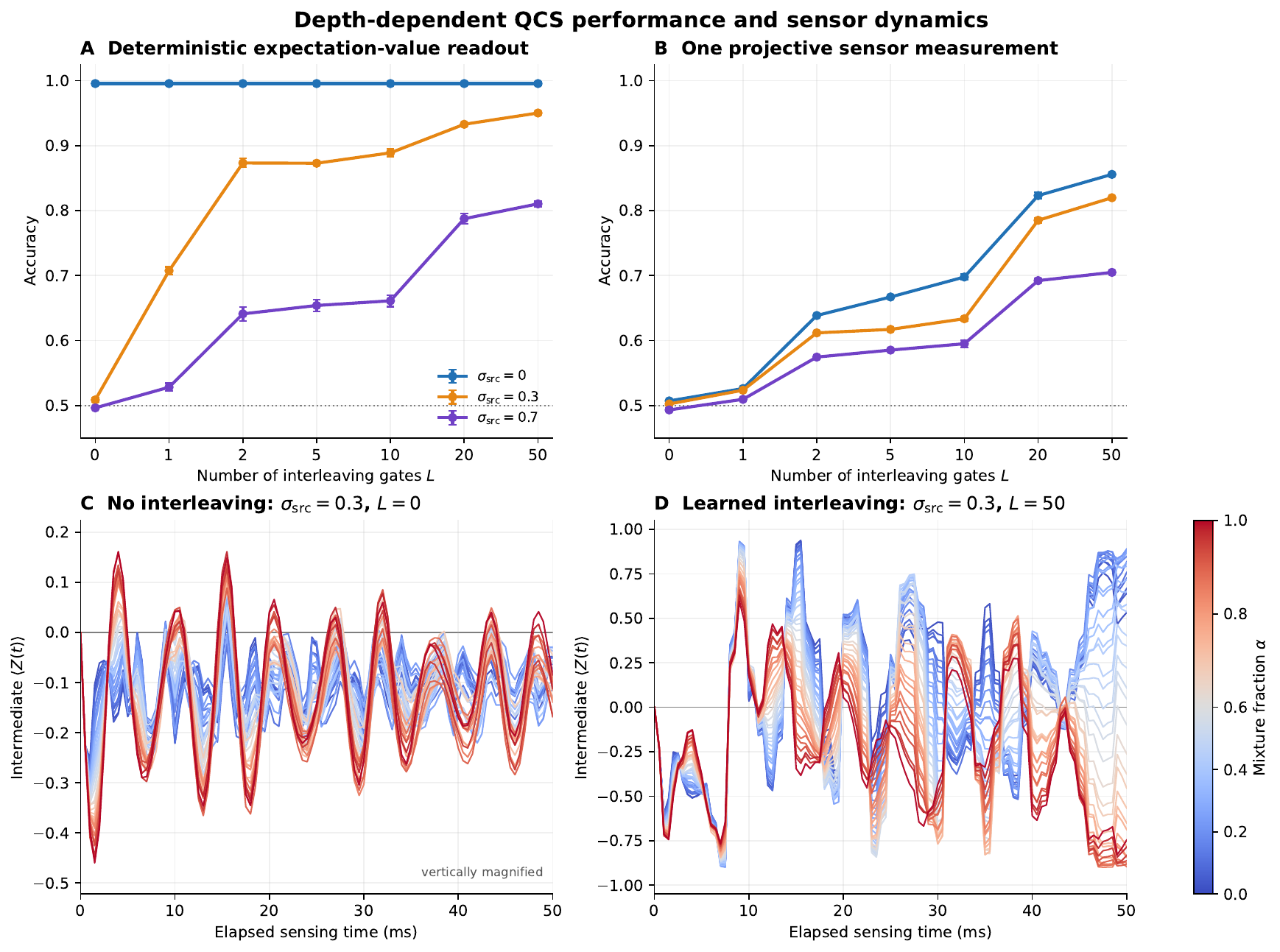}
\caption{
\textbf{Effect of interleaving depth on QCS performance and sensor
dynamics.}
Results use $g_{\rm sensor}=300~\mathrm{rad\,s^{-1}}$,
$T=T_2=50$ ms, $dt=0.5$ ms, and the source-noise levels indicated in the panels.
\textbf{A}, deterministic accuracy from the final expectation value
versus interleaving depth $L$.
\textbf{B}, expected correctness of one projective readout.
\textbf{C,D}, sensor trajectories averaged over independent
source-noise realizations at $\sigma_{\rm src}=0.3$, for
representative mixture fractions $\alpha$ indicated by color.
Panel C shows the bare sensor ($L=0$), read out at the
50 ms deadline, on a magnified vertical scale. Panel D
shows one representative trained $L=50$ control sequence;
the controls are held fixed while averaging over source noise.
}
\label{fig:depth_saturation_main}
\end{figure*}

\subsection{Effect of interleaving depth on QCS}
\label{sec:results-depth}

We first ask whether controls turn the available waveform
difference into a useful final response. For the main molecular
pair, the noiseless bare-sensor expectation is already separable
by a calibrated threshold at the 50 ms deadline. Source noise
degrades this shallow-protocol result; additional controls improve
deterministic accuracy
(Fig.~\ref{fig:depth_saturation_main}A).

One-shot correctness improves more gradually
(Fig.~\ref{fig:depth_saturation_main}B). A threshold can distinguish
small exact expectations, but a single binary outcome is informative
only when its probabilities differ sufficiently between classes.
The trajectories in panels C and D illustrate the corresponding
change in physical response: learned interleaving produces more
strongly mixture-dependent final expectations. These
noise-averaged trajectories illustrate the dynamics; accuracy is
evaluated on individual test examples.

We note that
the $L=0$ curve uses readout at the fixed 50 ms deadline, not the
bare sensor's best readout time. At $\sigma_{\rm src}=0.3$,
selecting an earlier time on validation data raises its held-out
deterministic accuracy to approximately $0.83$, while expected
one-shot correctness remains approximately $0.54$.
A separate, short-time
matched-readout comparison in that appendix shows that newly
trained controls also improve one-shot correctness over $L=0$
when both protocols end at the same selected time (Appendix~\ref{app:readout-timing}).

\subsection{Coupling and sensing duration under source noise}
\label{sec:results-duration}

\begin{figure*}[!t]
\centering
\includegraphics[width=0.95\linewidth]
{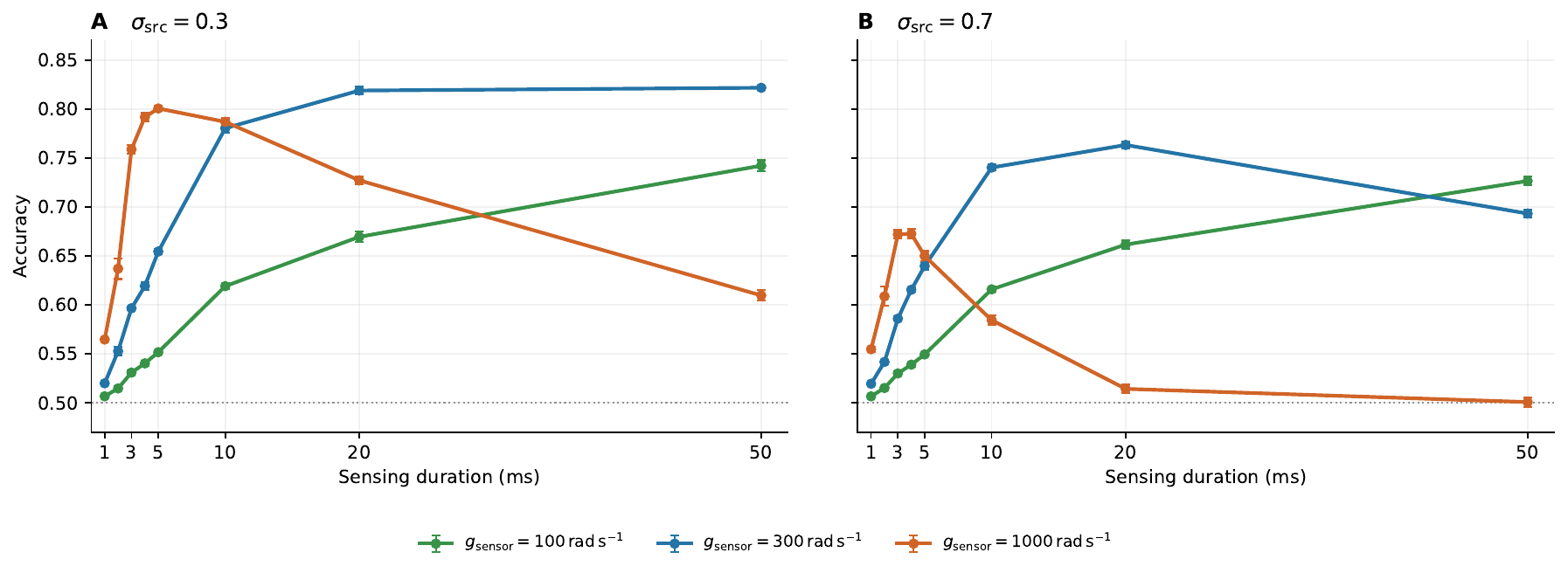}
\caption{
\textbf{Coupling--duration trade-off under source-waveform noise.}
Expected correctness of one projective sensor readout versus sensing duration for
$g_{\rm sensor}=100,\ 300,\ 1000~\mathrm{rad\,s^{-1}}$.
Panels \textbf{A} and \textbf{B} use per-sample, per-quadrature
source noise $\sigma_{\rm src}=0.3$ and $0.7$, respectively.
Each point evaluates a separately trained protocol on the same
molecular pair.
Each protocol uses the initial $T$ of the same pair-normalized molecular transient.
The sampling interval is $dt=0.5$ ms, so the quoted
per-sample noise levels correspond to
$S_{\rm src}=\sigma_{\rm src}^{\,2}dt$ per quadrature.
Two waveform steps form each 1 ms control segment; the plotted
durations $T=1,3,5,10,20,50$ ms therefore use depths
$L=1,3,5,10,20,50$, respectively.
Decision thresholds are calibrated on validation data.
Plotted accuracies are evaluated on separate held-out test data
and averaged over ten training seeds; error bars indicate
$\pm$SEM across seeds.
}
\label{fig:sensing_duration_three_couplings}
\end{figure*}

The action and noise-exposure scales of Sec.~\ref{sec:analytics-qcs}
suggest that stronger coupling need not favor a longer interaction.
Fig.~\ref{fig:sensing_duration_three_couplings} examines
expected correctness from one projective readout as coupling
and sensing duration vary. Each duration uses a separately
trained protocol, with a fixed 1 ms control-segment duration;
thus, increasing the sensing time also increases the control
depth $L$.

For example, at $\sigma_{\rm src}=0.3$ and
$g_{\rm sensor}=1000~\mathrm{rad\,s^{-1}}$, mean one-shot
correctness is $0.800$ at 5 ms but $0.610$ at 50 ms. At the
same noise level, $g_{\rm sensor}=300~\mathrm{rad\,s^{-1}}$
instead reaches $0.822$ at 50 ms. Increasing coupling can therefore
supply useful signal action sooner without making a longer
interaction advantageous.

At fixed normalized waveform and $\sigma_{\rm src}$,
increasing $g_{\rm sensor}$ scales both the coherent signal
and the source fluctuations driving the sensor. Available
signal action grows linearly with $g_{\rm sensor}$, whereas
the modeled source-noise exposure rate
$\Gamma_{\rm src}=g_{\rm sensor}^{\,2}S_{\rm src}$ grows
quadratically. This is consistent with the observed
coupling-duration trade-off, but does not determine an
optimum from $\Gamma_{\rm src}$ alone: changing duration
also changes the waveform portion sampled and the number
of controls. Appendix~\ref{app:coupling-duration} shows
that shortening the high-coupling protocol remains
beneficial when the control spacing is changed.

\subsection{Coupling and repeated acquisitions}
\label{sec:results-regimemap}

\begin{figure*}[!t]
\centering
\includegraphics[width=0.99\linewidth]
{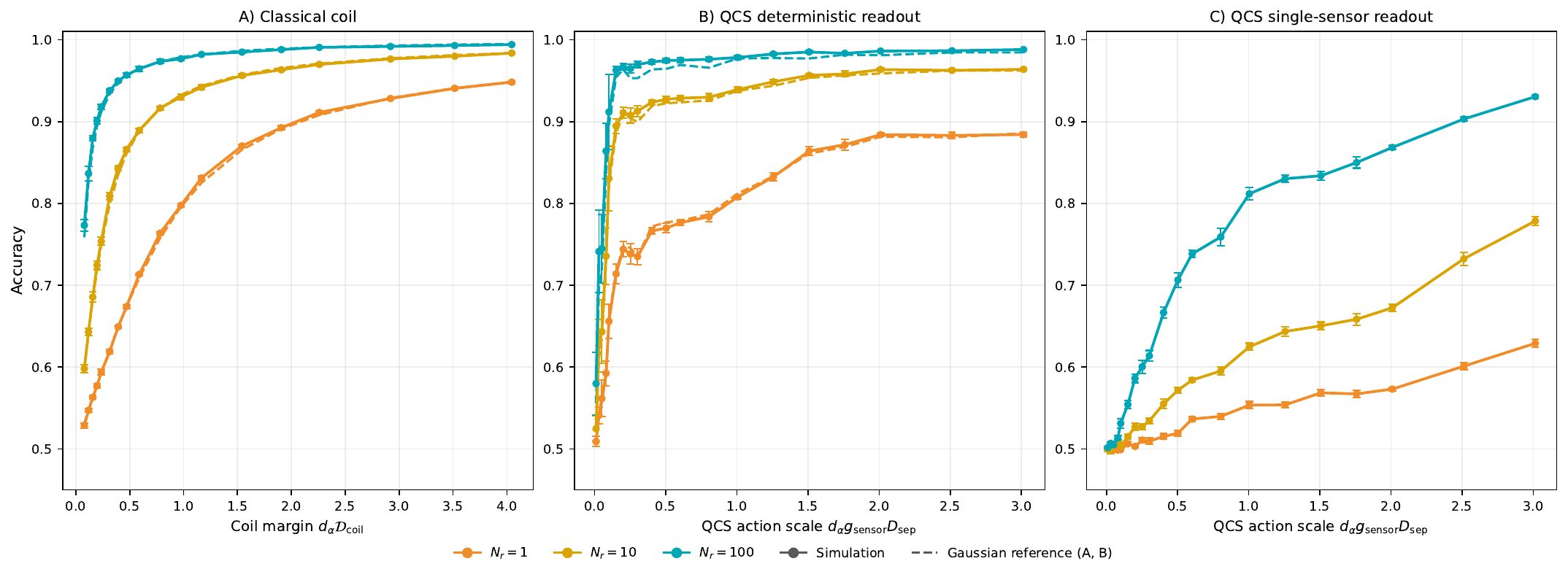}
\caption{
\textbf{Coupling and repeated acquisitions for coil and QCS readout.}
All panels use the same molecular pair,
$\sigma_{\rm src}=0.3$, $T=T_2=50$~ms, $dt=0.5$~ms, and
$N_r\in\{1,10,100\}$ independent acquisitions.
The coil has receiver noise $\eta_{\rm coil}=0.3$;
QCS uses control depth $L=10$.
At each sensor coupling, the trained QCS controls are held fixed
as $N_r$ varies; decision rules are calibrated separately for each $N_r$.
Solid curves show simulated accuracy.
Dashed curves show the Gaussian coil-model prediction (A)
and a Gaussian score-moment calculation using independently estimated
single-acquisition responses of the trained sensor (B).
Panel C shows one projective sensor outcome per acquisition,
classified using a validation-calibrated integer-count rule.
Error bars on the solid curves indicate $\pm$SEM over
independent simulation seeds; for QCS these seeds also vary
the training.
Here $d_\alpha=\mathbb E_\alpha|\alpha-\alpha_0|$
under the evaluation mixture distribution. Panel A uses
the coil margin $d_\alpha\mathcal D_{\rm coil}$; panels
B and C use the QCS action scale
$d_\alpha g_{\rm sensor}D_{\rm sep}$.
}
\label{fig:regime_map_main}
\end{figure*}

Fig.~\ref{fig:regime_map_main} shows how independent
acquisitions affect classification as the coil transfer
coefficient $g_{\rm coil}$ (panel A) or the effective sensor
drive strength $g_{\rm sensor}$ (panels B and C) is varied.
For the coil, increasing $g_{\rm coil}$ initially improves
accuracy against receiver noise. At strong transfer,
propagated source noise limits further improvement, while
repetition reduces both noise contributions. The dashed
curves average the Gaussian matched-filter prediction over
the full mixture distribution and closely follow the
simulations without fitting to their accuracies. Unlike the
simulated classifier, this prediction places its threshold
at the physical mixture boundary.

For deterministic QCS (panel B), each acquisition supplies the
exact final expectation for a newly drawn noisy waveform. The
solid curves classify averages of $N_r$ such expectations.
The dashed curves instead use independently estimated
single-acquisition means $\mu_\alpha$ and variances $v_\alpha$
at each mixture fraction. They approximate the acquisition
average as Gaussian with mean $\mu_\alpha$ and variance
$v_\alpha/N_r$, then apply the decision rule calibrated for
that $N_r$. Their agreement with the simulations shows that
this score-level approximation captures much of the repetition
dependence; small discrepancies can also arise from
non-Gaussian scores and finite sampling.
Small non-monotonic changes in the QCS curves may reflect
different local optima reached by training at each sensor-drive
strength, as well as finite-seed uncertainty; they should not
be taken as evidence of a physical trend on their own.

Panel C uses one projective outcome per independently noisy
acquisition rather than an exact expectation. Repetition
therefore averages both variation between source-waveform
realizations and projective-readout randomness, as described
by Eq.~\eqref{eq:Nr-Ns-qcs-main}. This explains why many
single-sensor acquisitions can outperform exact-expectation
readout of just one noisy transient: the former average over
many source-noise realizations, whereas the latter use only one realization.

The horizontal coordinates include the molecular waveform
difference and the mean distance from the mixture boundary.
They have different meanings: $d_\alpha\mathcal D_{\rm coil}$
is a noise-normalized coil margin, whereas
$d_\alpha g_{\rm sensor}D_{\rm sep}$ is the mean scale
of an upper bound on physical expectation differences
relative to the boundary mixture. It is not the contrast
actually achieved by the trained sensor. Accuracy in all
panels is evaluated over the full mixture distribution,
not only at its mean boundary distance.

\subsection{Mitigating quasistatic sensor-splitting disorder}
\label{sec:results-quasistatic}

\begin{figure}[!t]
\centering
\includegraphics[width=\linewidth]
{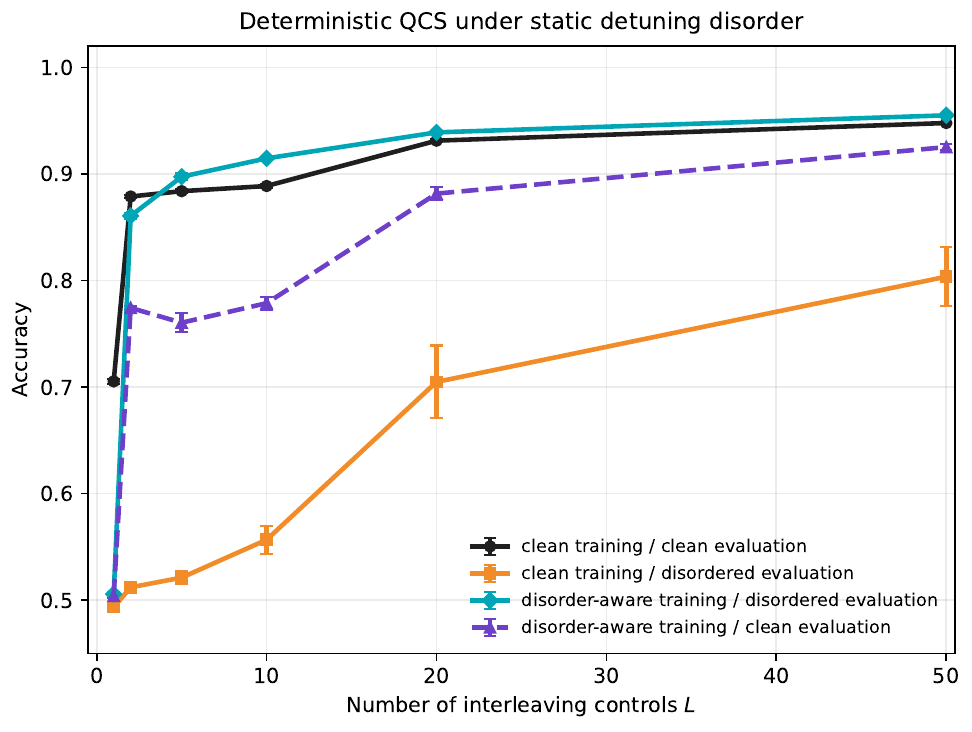}
\caption{
\textbf{Disorder-aware training mitigates quasistatic
sensor-splitting disorder in ensemble QCS.}
A static term $H_{\rm dis}=\Delta Z/2$, with
$\Delta\sim\mathcal N(0,\sigma_\Delta^2)$ and
$\sigma_\Delta=300~\mathrm{rad\,s^{-1}}$, acts during sensing.
Clean- and disorder-aware-trained protocols are evaluated
on clean or detuning-averaged sensor responses as indicated.
Each protocol applies the same instantaneous control gates
to every sensor in the ensemble. Results are shown versus
control depth $L$ at
$g_{\rm sensor}=300~\mathrm{rad\,s^{-1}}$ and
$\sigma_{\rm src}=0.3$. Curves show means over independent
training seeds; error bars indicate $\pm$SEM.
}
\label{fig:disorder_robustness_main}
\end{figure}

Finally, we consider a sensor limitation distinct from the
source-waveform fluctuations above. Static spatial variation
in sensor splitting can reduce the expectation averaged over
an ensemble. We model this by adding
\begin{equation}
H_{\rm dis}=\frac{\Delta}{2}Z,
\qquad
\Delta\sim\mathcal N(0,\sigma_\Delta^2),
\end{equation}
during sensing and classifying the detuning-averaged response.

Fig.~\ref{fig:disorder_robustness_main} shows that clean-trained
controls can lose accuracy under disorder. Training a single,
globally applied protocol on the disorder-averaged response
instead restores high ensemble accuracy across the tested
controlled depths. Its mean accuracy occasionally exceeds the
clean-reference mean slightly, although the figure does not
establish a systematic advantage over that reference. At
smaller depths, disorder-aware controls can also perform less
well when evaluated on a clean ensemble.

This recovery concerns the \emph{ensemble-averaged} score;
individual detuning slices need not all classify well. The
model assumes instantaneous gates that act identically across
sensors. With finite-duration gates, detuning can also affect
the gates themselves; their speed or robustness would need
to be included in an implementation-specific analysis.


\section{Conclusions}
\label{sec:conclusion}

We studied task-oriented classification of transient
molecular-NMR signals using inductive waveform readout and
quantum computational sensing (QCS). For the coil route,
we derive a Gaussian accuracy relation: classification
depends on the difference between the pure-component
waveforms, the noise level, and how close a mixture is to
the decision threshold.
For QCS, the mixture-dependent waveform difference bounds
the possible sensor contrast, while trained controls
determine the response actually achieved.
Interleaving can improve both classification from exact
expectation values and the contrast needed for an informative
projective readout.

Effective drive strength and sensing duration must be
considered together. At fixed normalized waveform and
source-noise level, increasing $g_{\rm sensor}$ scales both
the coherent signal and the source fluctuations reaching
the sensor. The coherent drive grows linearly, while the
source-noise exposure rate
$\Gamma_{\rm src}=g_{\rm sensor}^2S_{\rm src}$ grows
quadratically. Different source-noise realizations produce
different final sensor responses: averaging them can
attenuate the mean response, while their variation limits
classification from individual expectation values. These
are two effects of the same source noise, not independent
penalties.
At strong effective drive, the numerical results show that
shorter, separately trained QCS protocols can outperform
longer ones in projective-readout classification.
Intrinsic sensor decoherence, not simulated
here, may further restrict the usable sensing window.
Neither the source-noise rate nor the sensor coherence time
alone determines the optimal duration.

The numerical results support the analytical description of
repeated acquisitions. Independent repetition reduces
source-induced fluctuations in both routes, receiver noise
in the coil, and projection noise in finite-sensor QCS.
The Gaussian coil model closely describes the simulated coil
results, while single-acquisition QCS score statistics capture
much of the observed dependence on repetition. Separately,
disorder-aware control mitigates quasistatic splitting
inhomogeneity in the modeled sensor ensemble.

Practical implementation remains challenging. In the
high-field NMR mixture benchmark, strong one-shot performance
at $g_{\rm sensor}=300~\mathrm{rad\,s^{-1}}$ occurs for
total sensing durations $T$ of order tens of milliseconds.
The sensor must retain a useful signal response throughout
the coherent signal-and-control sequence, without
intermediate measurement or reset. The separate 1~ms study
in Appendix~\ref{app:readout-timing} instead uses
$g_{\rm sensor}=10^4~\mathrm{rad\,s^{-1}}$, a much stronger
effective drive. For a constant, unit-amplitude resonant
envelope, $g_{\rm sensor}$ is the qubit's angular rotation
rate in our Hamiltonian convention. Under the illustrative
NV-electron conversion in
Appendix~\ref{app:parameter_interpretation}, these drive
strengths have field equivalents of about $1.7$ and $57$ nT,
respectively. The former is comparable in order of magnitude
to NMR fields reported in microscale NV experiments~\cite{Smits2019MicrofluidicDiamond}.
However, an NV implementation of the high-field NMR benchmark requires
a specified signal-to-sensor interface.

The lower-field library benchmark at $B_0=0.1~\mathrm{T}$
(Appendix~\ref{app:real_library_identification}) may offer
a more accessible starting point for exploring an NV-based
interface. Its $40$~s total sensing duration, however, poses
a demanding requirement for continuous, signal-sensitive
sensor evolution. Measuring and resetting sensors between
shorter segments could relax that coherence requirement~\cite{Khan2026QCS},
but would constitute a different protocol whose classification performance and readout costs
must be assessed separately. Identifying a suitable platform
and assessing its signal transfer, achievable drive strength,
coherence under compatible controls, and readout remain
important directions for future work.


\section*{Acknowledgements}

We thank Gary Schmiedinghoff, Johannes Renkl, and Peter Schuhmacher for useful discussions.
This work was supported by the German Federal Ministry of
Research, Technology and Space (BMFTR) through the project
``Algorithms for quantum computer development in
hardware-software codesign'' (ALQU), made possible by the
DLR Quantum Computing Initiative (QCI). Additional support
was provided by the BMFTR through the project
``Neuronale Quantennetzwerke auf NISQ-Quantencomputern''
(NeuQuant), grant no.~13N17067.

Artificial-intelligence tools, including ChatGPT (GPT-5.2--GPT-6) and
Anthropic Claude (Sonnet 4.6 and Sonnet 5),
assisted with manuscript drafting, code development,
and analytical checks. The authors directed their use and
take full responsibility for the work.


\appendix

\section{Coil classification: Gaussian matched-filter calculation}
\label{app:coil-analytic-derivation}

We derive the coil reference used in
Sec.~\ref{sec:analytics-coil} and
Fig.~\ref{fig:regime_map_main}. Let
$\bar{\mathbf s}_j=(s_k^{(j)}e^{-t_k/T_2})_{k=0}^{K-1}$,
$\mathbf v=\bar{\mathbf s}_A-\bar{\mathbf s}_B$, and
$\delta=\alpha-\alpha_0$. The sampled observation can be written
\begin{equation}
\mathbf c
=g_{\rm coil}(\bar{\mathbf s}_{\alpha_0}+\delta\mathbf v)
+\mathbf n,\qquad
\mathbf n=g_{\rm coil}\boldsymbol\xi_{\rm src}
+\boldsymbol\xi_{\rm coil}.
\label{eq:app-coil-vector}
\end{equation}
The real and imaginary components of $\mathbf n$ are independent
Gaussians of variance
$\sigma_{\rm eff}^2=g_{\rm coil}^2\sigma_{\rm src}^2+
\eta_{\rm coil}^2$. We use the complex inner product
$\langle\mathbf a,\mathbf b\rangle=\sum_k a_k^*b_k$.
For a score centered at the physical boundary, define
\begin{equation}
S_{\rm coil}
=\operatorname{Re}
\langle\mathbf c-g_{\rm coil}\bar{\mathbf s}_{\alpha_0},
\mathbf v\rangle.
\label{eq:coil-score-main}
\end{equation}
The projected noise has variance
$\sigma_{\rm eff}^2\|\mathbf v\|^2$, so
\begin{equation}
S_{\rm coil}\mid\delta
\sim\mathcal N\!\left(
g_{\rm coil}\delta\|\mathbf v\|^2,\,
\sigma_{\rm eff}^2\|\mathbf v\|^2\right).
\label{eq:app-coil-score-distribution}
\end{equation}
Thresholding at zero gives the main-text expression
$P_{\rm corr}^{\rm coil}(\delta)
=\Phi(|\delta|\mathcal D_{\rm coil})$,
where $\mathcal D_{\rm coil}=
g_{\rm coil}\|\mathbf v\|/\sigma_{\rm eff}$.
For $N_r$ independent waveforms averaged before classification,
the score variance decreases by $N_r$; the ideal accuracy over
the prescribed evaluation distribution $p_{\rm eval}(\alpha)$ is
\begin{equation}
\operatorname{Acc}_{\rm coil}^{\rm ideal}(N_r)
=\mathbb E_{\alpha\sim p_{\rm eval}}
\!\left[
\Phi\!\left(
\sqrt{N_r}\,|\alpha-\alpha_0|\mathcal D_{\rm coil}
\right)\right].
\label{eq:app-coil-ideal-accuracy}
\end{equation}
This is the parameter-free dashed reference in
Fig.~\ref{fig:regime_map_main}A; the expectation is over the
\emph{full} mixture distribution, not its mean boundary distance.

The numerical classifier uses a linear template-difference
direction parallel to $\mathbf v$, but selects a scalar threshold
on separate calibration data. For the centered score above and a
threshold $h_{N_r}$, its Gaussian conditional correctness is
\begin{equation}
P_{\rm corr}^{\rm coil}(\delta;h_{N_r})
=
\Phi\!\left[
\frac{
y(\delta)\sqrt{N_r}
(g_{\rm coil}\delta\|\mathbf v\|^2-h_{N_r})
}{
\sigma_{\rm eff}\|\mathbf v\|}
\right].
\label{eq:app-coil-calibrated-accuracy}
\end{equation}
Here $y(\delta)=+1$ for $\delta\geq0$ and
$y(\delta)=-1$ for $\delta<0$, as in
Eq.~\eqref{eq:mixture-label}.
A different normalization of the linear score rescales
$h_{N_r}$ accordingly. The ideal reference sets $h_{N_r}=0$;
the plotted simulation uses its calibrated rule.

\section{Molecular benchmark and numerical procedures}
\label{app:simulation-protocol}

\subsection{Molecular waveforms and normalization}

The molecular inputs are proton-only spin systems from the HQS NMR
parameter collections~\cite{hqs_nmr_parameters}. Each supplies
chemical shifts $\delta_i$ (ppm) and scalar couplings $J_{ij}$
(Hz). We use the Hamiltonian of
Eq.~\eqref{eq:molecular-hamiltonian}, with
$\omega_i=2\pi\nu_i$,
$J_{ij}^{(\omega)}=2\pi J_{ij}$, and
\begin{equation}
\nu_i=\nu_0(\delta_i-\delta_{\rm ref})10^{-6},
\qquad
\nu_0=\left(\frac{\gamma_{^1{\rm H}}}{2\pi}\right)B_0.
\label{eq:app-frequency-conversion}
\end{equation}
The main pair consists of imidazole
(\texttt{gissmo-experimental/bmse000096}) and isoxazole
(\texttt{cheshire-combined/30}). Mixture comparisons use a
\emph{shared}, pair-centered rotating frame:
$\delta_{\rm ref}=(\bar\delta_A+\bar\delta_B)/2$, where
$\bar\delta_j$ is molecule $j$'s mean proton shift. Thus their
relative offsets are retained.

The default waveform is the collective transverse signal following
initialization of the molecular spins in
$|+\rangle^{\otimes n}$:
\begin{equation}
s(t)=I(t)+iQ(t),\qquad
|+\rangle=(|0\rangle+|1\rangle)/\sqrt2.
\label{eq:app-molecular-waveform}
\end{equation}
We also checked a high-temperature FID-inspired preparation:
$\rho_{\rm dev}(0)=n^{-1}\sum_i X_i$, propagated under
$H_{\rm mol}$, with~\cite{Fratus2025NMRQuantumSimulation}
\begin{equation}
s(t)=2^{-n}\operatorname{Tr}
[\rho_{\rm dev}(t)(X_{\rm tot}+iY_{\rm tot})].
\label{eq:app-thermal-fid}
\end{equation}
The reported main benchmark uses the product-state preparation;
the second preparation is a model check, not a distinct plotted
benchmark. In either case, the clean signal is multiplied by
$e^{-t/T_2}$ before the independent source noise is added.

For each mixture pair, the generator first applies the
phenomenological $T_2$ envelope to both molecular waveforms.
The resulting noise-free waveforms are divided by one common
early-time reference amplitude. With $r_A$ and $r_B$ their
respective RMS amplitudes over the first $K_{\rm ref}=10$
samples, this amplitude is
\begin{equation}
r_{\rm ref}=\tfrac12(r_A+r_B).
\label{eq:pair-common-normalization}
\end{equation}
The common normalization preserves relative amplitudes within
the pair; source noise is added afterward. The main mixture
benchmark uses $B_0=11.74$ T, $K=100$, $dt=0.5$ ms,
$T=50$ ms, $T_2=50$ ms, and $\alpha_0=0.6$.
Source-noise amplitudes $\sigma_{\rm src}$ are specified per
sampled time step and per quadrature; changing $dt$ at fixed
$\sigma_{\rm src}$ changes the effective continuous-time
intensity $S_{\rm src}=\sigma_{\rm src}^2dt$.

In the main-text notation, $s^{(m)}(t)$ denotes the
undecayed molecular waveform divided by this reference;
the noise-free array used in the simulations is therefore
$s^{(m)}(t_k)e^{-t_k/T_2}$.

\subsection{QCS controls and evaluation}
\label{app:qcs-optimizer}

For depth $L$, the trainable controls are
$V_\ell=R_z(\phi_\ell)R_x(\theta_\ell)$,
$\ell=1,\ldots,L$. In expectation-value training the
physical final expectation is combined with a trainable affine
decision head; we denote its pre-activation by $u=wz+b$ to
distinguish it from the physical expectation $z=\langle Z\rangle$.
One training objective used in the deterministic benchmarks is
\begin{equation}
\mathcal L_{\rm exp}
=\frac1B\sum_{i=1}^{B}
\left[
(\tanh u_i-y_i)^2+
\lambda_{\rm log}\operatorname{softplus}(-y_i u_i)
\right],
\label{eq:appendix-exp-loss}
\end{equation}
with $\lambda_{\rm log}=0.2$.
The principal repetition comparison keeps trained controls fixed
while varying $N_r$; it calibrates the decision rule separately
for each $N_r$. For its $N_s=1$ panel, that rule is a
validation-calibrated boundary on the integer number of positive
outcomes.

All QCS protocols reported in this paper are trained with the
expectation-value objective in Eq.~\eqref{eq:appendix-exp-loss}.
Expected one-shot correctness is evaluated from projective
readout of these protocols; it is not the training objective.
Protocols described as separately trained at different sensing
durations use the same expectation-value objective.

Training uses Adam~\cite{Kingma2015Adam} with a cosine
learning-rate schedule~\cite{Loshchilov2017SGDR}.
Here $\eta_{\max}$ and $\eta_{\min}$ denote the initial
and final learning rates. The default setting uses
160 epochs, with $\eta_{\max}=0.05$ and
$\eta_{\min}=0.2\eta_{\max}$. Independent simulation seeds vary the initial trainable
parameters, sampled training data, and source-noise
realizations. Plotted QCS points are means of the
held-out accuracies obtained across these seeds; error
bars show the standard error of that mean (SEM), not the
full seed-to-seed spread. The results represent trained
performance, not certified global optima.

The dashed QCS calculation in Fig.~\ref{fig:regime_map_main}B
uses fresh single-acquisition waveforms, independently of the
plotted test outcomes, to estimate
$\mu_\alpha=\mathbb E_\xi[z_\xi(\alpha)]$ and
$v_\alpha=\operatorname{Var}_\xi[z_\xi(\alpha)]$ at each
mixture fraction, coupling, and training seed. It applies the
corresponding validation-calibrated rule to a Gaussian with
mean $\mu_\alpha$ and variance $v_\alpha/N_r$, then
averages accuracy over mixtures and seeds. It does not fit the
plotted test accuracy; Gaussianity of the averaged score is
the approximation being checked.

\begin{figure*}[!t]
\centering
\includegraphics[width=0.98\linewidth]
{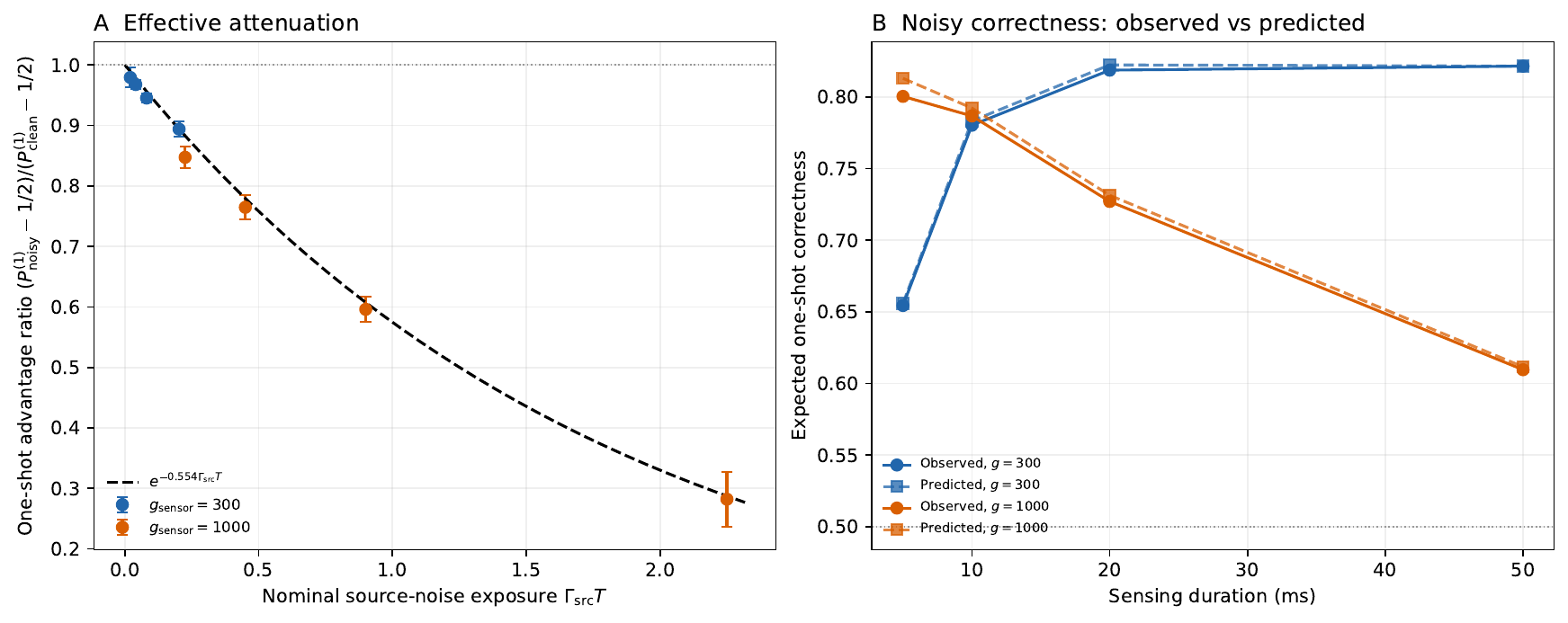}
\caption{
\textbf{Empirical attenuation of one-shot advantage under
source noise.}
For each protocol, clean and noisy evaluations use the same
trained controls and decision rule, calibrated on noisy
validation data. ``Clean'' means that additive source noise
is set to zero; the $T_2$ envelope is retained.
\textbf{A}, ratio of noisy to clean advantages over chance
versus $\Gamma_{\rm src}T$ at $\sigma_{\rm src}=0.3$.
The dashed curve is $e^{-c\Gamma_{\rm src}T}$, with
$c=0.554$ calibrated only at
$g_{\rm sensor}=300~\mathrm{rad\,s^{-1}}$ and $T=50$ ms.
\textbf{B}, directly evaluated noisy one-shot correctness
and values calculated from clean correctness using the
same $c$. Durations are $5,10,20,50$ ms, with 1 ms between
control-segment boundaries. Points summarize ten training
seeds; error bars in A show the seed-to-seed standard
deviation of the advantage ratio. The relation describes
aggregate one-shot correctness, not deterministic accuracy.
}
\label{fig:app-source-attenuation}
\end{figure*}

\section{Source-noise-averaged sensor dynamics}
\label{app:source-noise-channel}

The simulated source noise is redrawn independently in the two
signal quadratures at every sampling step. Conditional on an
individual waveform, sensor evolution is unitary; averaging
over waveforms produces a nonunitary mean-state channel.

For a step of length $dt$, write
\begin{equation}
H_k=H_{{\rm clean},k}
+\frac{g_{\rm sensor}}{2}
(\xi_{I,k}X+\xi_{Q,k}Y),
\label{eq:app-source-noise-H}
\end{equation}
where the independent noise samples satisfy
$\mathbb E[\xi_{a,k}]=0$ and
$\mathbb E[\xi_{a,k}^{\,2}]=\sigma_{\rm src}^{\,2}$
for $a\in\{I,Q\}$.
Expanding the step propagator through second order in the
noise-driven rotation and averaging gives
\begin{align}
\bar\rho_{k+1}-\bar\rho_k &\simeq-i\,dt[H_{{\rm clean},k},\bar\rho_k] \nonumber \\
&-\frac{g_{\rm sensor}^2\sigma_{\rm src}^2dt^2}{8}
\sum_{a=X,Y}[a,[a,\bar\rho_k]].
\label{eq:app-source-noise-step}
\end{align}
Dividing by $dt$ yields the main-text master equation with
$\Gamma_{\rm src}=g_{\rm sensor}^2\sigma_{\rm src}^2dt$.
The control gates act at their prescribed boundaries. With no
clean drive or controls, the mean Bloch coordinates decay as
$e^{-\Gamma_{\rm src}t/2}$ for $x,y$ and
$e^{-\Gamma_{\rm src}t}$ for $z$.
During sensing, the molecular signal and controls mix the
components: neither rate by itself predicts a classification
accuracy or an optimal sensing time.

In particular, the averaged channel yields
$\mu_\alpha=\operatorname{Tr}(Z\bar\rho_\alpha)$, but not the
distribution of individual noisy-waveform expectations.
The variance $v_\alpha$ required for deterministic
single-acquisition accuracy depends on the second moment
$\mathbb E_\xi[z_\xi(\alpha)^2]$. The attenuation of a mean
response and the dispersion of conditioned scores arise from
the \emph{same} source fluctuations; they must not be counted
as separate, independent penalties in the local Gaussian
correctness formula.

For one projective readout, the situation is different:
with a fixed outcome-to-label rule, expected correctness is
linear in the waveform-conditioned expectation
$z_\xi(\alpha)$. Averaging over source noise therefore
requires only its mean $\mu_\alpha$, not the full
distribution of individual scores. The projective-outcome
probabilities account for measurement randomness.

Motivated by the attenuation of the averaged sensor response,
we also test a compact empirical description of one-shot
correctness. For each saved
protocol, we compare expected correctness with the modeled
source noise present and with the source noise set to zero.
Both evaluations use the same trained controls and
the decision rule calibrated on noisy validation data:
\begin{equation}
P_{\rm noisy}^{(1)}-\frac12
\approx
e^{-c\Gamma_{\rm src}T}
\left(P_{\rm clean}^{(1)}-\frac12\right).
\label{eq:app-source-noise-attenuation}
\end{equation}
Here $c=0.554$ is fixed using one operating point, rather
than derived from the master equation. As shown in
Fig.~\ref{fig:app-source-attenuation}, it describes the
remaining tested protocols reasonably well.

\begin{figure*}[!t]
\centering
\includegraphics[width=0.98\linewidth]
{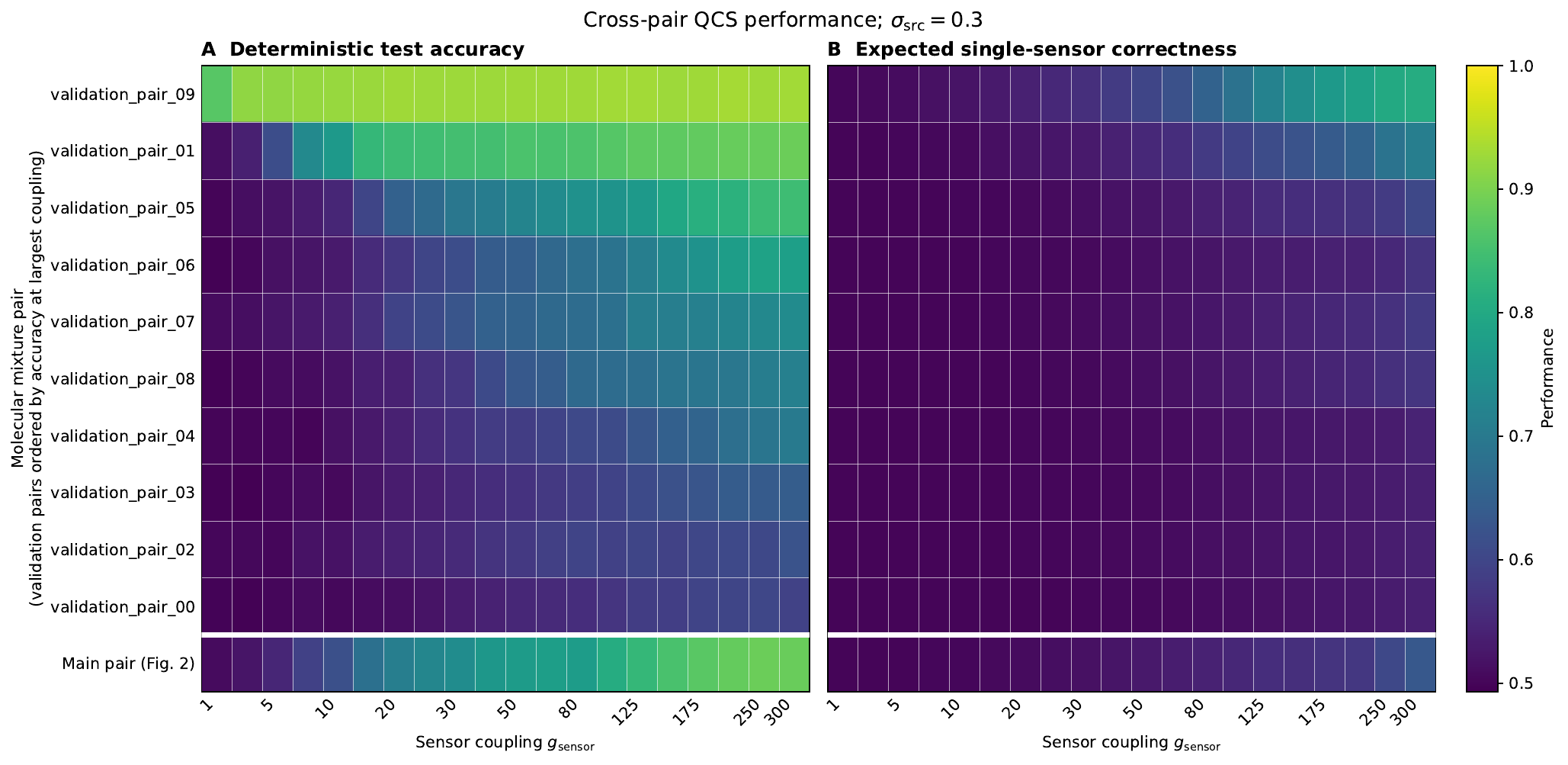}
\caption{
\textbf{QCS performance across molecular mixture pairs.}
\textbf{A}, deterministic held-out accuracy;
\textbf{B}, expected correctness from one projective readout
of the expectation-value-trained sensor.
All results use $L=10$, $T=T_2=50$ ms, $dt=0.5$ ms,
and $\sigma_{\rm src}=0.3$; each cell averages ten training
seeds. For display, the ten additional pairs are ordered
by their mean deterministic accuracy at
$g_{\rm sensor}=300~\mathrm{rad\,s^{-1}}$.
The main-text pair is shown separately at the bottom.
Columns show the tested sensor couplings; chance
performance is $0.5$.
}
\label{fig:cross_pair_performance}
\end{figure*}

\section{Additional molecular tasks}
\label{app:additional-tasks}

\subsection{Variation across mixture pairs}
\label{app:mixture_pair_robustness}

We selected ten additional molecular pairs spanning a range
of waveform similarities, without using their QCS results
to choose the pairs. Fig.~\ref{fig:cross_pair_performance}
shows results at control depth $L=10$.
For display, the validation pairs are ordered by their mean
deterministic QCS accuracy at
$g_{\rm sensor}=300~\mathrm{rad\,s^{-1}}$; the main-text pair
is shown separately at the bottom.

As Fig.~\ref{fig:cross_pair_performance} shows, both
performance and its coupling dependence vary appreciably
between pairs. Deterministic separability need not imply
informative one-shot outcomes. The figure establishes
pair dependence, not a universal performance ranking from
a simple waveform-distance measure.

\begin{figure*}[!t]
\centering
\includegraphics[width=0.98\linewidth]
{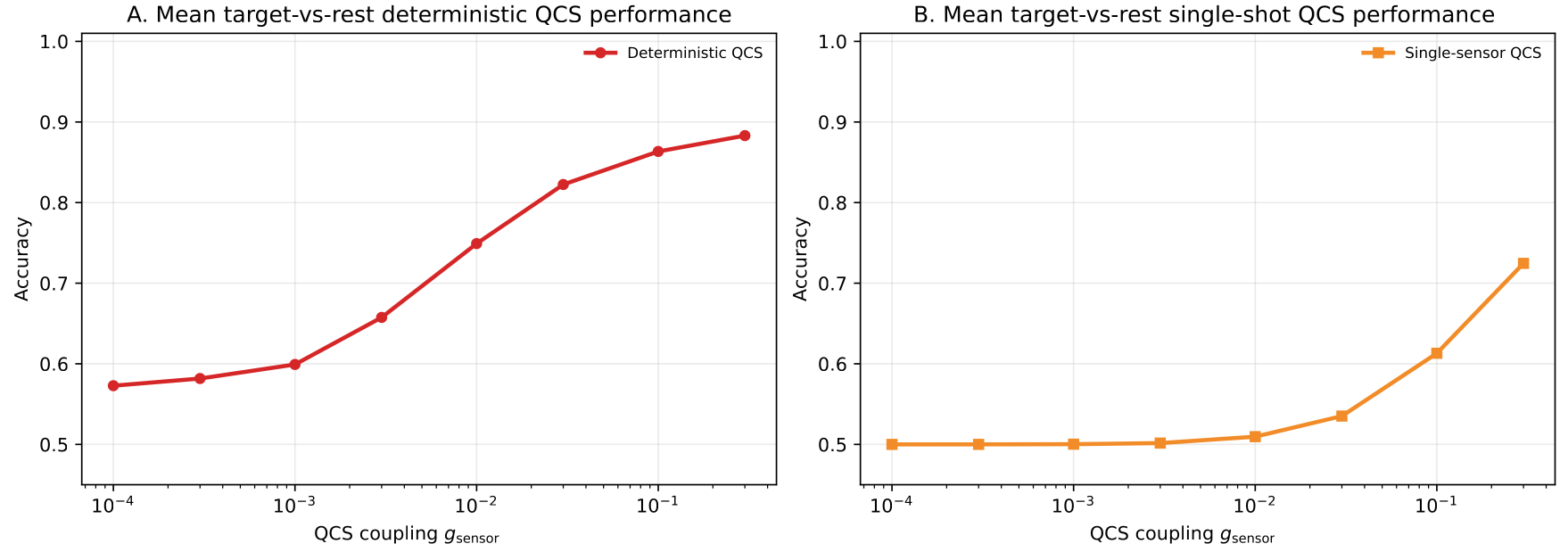}
\caption{
\textbf{Centered-frame molecular-library identification.}
A balanced target-vs-rest task is defined for each of 114
five-spin molecules, with each waveform generated in its own
mean-shift-centered rotating frame.
\textbf{A}, target-averaged deterministic QCS accuracy;
\textbf{B}, target-averaged expected correctness from one
projective readout of the expectation-value-trained sensor.
Both are shown versus the effective sensor-drive scale at
$B_0=0.1$ T, $K=100$, $dt=0.4$ s, $T=T_2=40$~s, $L=10$, $\sigma_{\rm src}=0.1$, and $N_r=1$.
The horizontal axes show the effective sensor-drive parameter
$g_{\rm sensor}$ at the tested values
$10^{-4},\,3\times10^{-4},\,10^{-3},\,3\times10^{-3},
\,10^{-2},\,3\times10^{-2},\,0.1,$ and
$0.3~\mathrm{rad\,s^{-1}}$.
}
\label{fig:real_library_identification_appendix}
\end{figure*}

\subsection{One-vs-rest molecular-library task}
\label{app:real_library_identification}

As a separate task, we classify whether an NMR signal comes
from a selected target molecule rather than another library
member. The library comprises 114 proton-only five-spin
molecules from the GISSMO, CHESHIRE, and HQS
collections~\cite{hqs_nmr_parameters}. For each target-vs-rest
task, evaluation assigns equal weight to target and non-target
signals.

In contrast to the shared rotating frame used for each mixture
pair, each library molecule is placed in its \emph{own}
mean-shift-centered rotating frame before waveform generation.
This is a centered-frame identifiability benchmark, not a
single common-frame deployment experiment. The library
benchmark uses $B_0=0.1$~T, $K=100$, $dt=0.4$~s,
$T=T_2=40$~s, and control depth $L=10$; each waveform is
normalized before source noise is added.

At $B_0=0.1~\mathrm{T}$, the proton carrier is approximately
$4.26~\mathrm{MHz}$. For an NV axis aligned with this field,
a simple estimate places its lower electronic transition near
$70~\mathrm{MHz}$. Compared with the high-field mixture
benchmark, this lower-frequency setting may offer more
accessible routes to frequency matching or conversion.

Fig.~\ref{fig:real_library_identification_appendix}
shows target-averaged performance at $\sigma_{\rm src}=0.1$, $dt=0.4$~s, and single acquisition~$N_r=1$.
Deterministic accuracy rises more rapidly with effective
sensor-drive strength than expected correctness from one
projective readout. The distinction between the two readout
regimes therefore persists in this separate molecular-library
task.

\section{Timing checks and physical parameters}
\label{app:supplementary-timing-parameters}

\subsection{Bare-sensor readout timing}
\label{app:readout-timing}

For the main pair at $g_{\rm sensor}=300~\mathrm{rad\,s^{-1}}$
and $\sigma_{\rm src}=0.3$, the trained $L=50$ protocol
at the 50 ms deadline reaches approximately $0.95$
deterministic accuracy and $0.82$ expected one-shot
correctness (Fig.~\ref{fig:depth_saturation_main}).
The bare sensor ($L=0$) achieves only $0.505$ and
$0.503$, respectively, when read at that deadline.
Allowing its readout time to be selected on validation
data raises held-out deterministic accuracy to $0.833$
at 1.5 ms, but expected one-shot correctness reaches
only $0.543$. The deterministic and one-shot criteria
select readout times separately; neither involves
intermediate measurement and reset.

A separate matched-time, one-shot check asks whether
controls add value beyond selecting an early readout.
It uses the original pair's amplitude normalization,
but a distinct grid of $K=100$, $dt=10~\mu\mathrm{s}$,
and $T=1$ ms. The stronger effective drive,
$g_{\rm sensor}=10^4~\mathrm{rad\,s^{-1}}$, compensates
in part for the much shorter time available to accumulate
signal action. The $L=0$ readout time is selected on
validation data, and a controlled protocol is newly
trained to end at that same time. This comparison tests
the benefit of controls at matched duration; it is a
separate coupling-duration regime from the 50~ms study.
\begin{table}[t]
\centering
\caption{
\textbf{Matched-time, one-projective-readout comparison.}
Held-out expected correctness averaged over five training
seeds. Each $L=0$ readout time $t_\star$ is selected
on validation data, and matched-time QCS is newly
trained to end there. Deadline QCS uses
the full 1~ms ($L=50$). Times and noise levels belong to
this separate 1 ms simulation grid.
}
\label{tab:matched-time-one-shot}
\begin{tabular}{ccccc}
\hline
$\sigma_{\rm src}$&
$t_\star$ (ms)&
$L=0$ at $t_\star$&
QCS at $t_\star$&
QCS at 1 ms\\
\hline
$0$ & $0.56$ & $0.737$ & $0.804$ & $0.878$\\
$1$ & $0.54$--$0.56$ & $0.653$ & $0.698$ & $0.728$\\
\hline
\end{tabular}
\end{table}
Table~\ref{tab:matched-time-one-shot} shows a
matched-time control benefit in the tested seeds.

\subsection{Coupling, duration, and control cadence}
\label{app:coupling-duration}

We first ask how much one-shot QCS performance improves
when total sensing duration~$T$ is selected rather than fixed at
50 ms. For each duration, a separate protocol is trained
on the initial portion of the same pair-normalized
transient. The waveform sampling interval is $dt=0.5$~ms,
and interleaving gates are spaced 1 ms apart. We select
$T$ on validation data from the predeclared
$5,10,20,50$~ms grid, corresponding to
$L=5,10,20,50$, and report held-out correctness in
Table~\ref{tab:coupling-duration-selection}.
Figure~\ref{fig:sensing_duration_three_couplings}
also shows shorter durations outside this selection grid.

\begin{table}[t]
\caption{\label{tab:coupling-duration-selection}
Validation-selected sensing durations. Each seed selects from
$T=5,10,20,50$~ms. Entries are means over ten seeds;
$T_{\rm sel}$ is the mean selected duration, and the two
accuracy columns give held-out one-shot correctness for the
selected and separately trained 50 ms protocols, respectively.
The coupling $g_{\rm sensor}$ is in $\mathrm{rad\,s^{-1}}$ and
$T_{\rm sel}$ is in ms.}
\centering
\setlength{\tabcolsep}{3pt}
\begin{tabular}{ccccc}
\hline
$\sigma_{\rm src}$ &
$g_{\rm sensor}$ &
$T_{\rm sel}$ &
$\mathrm{Acc}_{\rm sel}$ &
$\mathrm{Acc}_{50}$ \\
\hline
$0.3$ & $300$  & $41$  & $0.823$ & $0.822$ \\
$0.3$ & $1000$ & $5.5$ & $0.799$ & $0.610$ \\
$0.7$ & $300$  & $20$  & $0.763$ & $0.693$ \\
$0.7$ & $1000$ & $5$   & $0.650$ & $0.501$ \\
\hline
\end{tabular}
\end{table}

Duration selection has little effect at
$g_{\rm sensor}=300~\mathrm{rad\,s^{-1}}$ and
$\sigma_{\rm src}=0.3$, but improves performance in the
other tested cases (Table~\ref{tab:coupling-duration-selection}).

In an additional check not shown in the table or figure,
we repeated the duration comparison with 2.5 ms rather
than 1 ms between interleaving gates; the sampling interval
remained $dt=0.5$ ms. At
$g_{\rm sensor}=1000~\mathrm{rad\,s^{-1}}$ and
$\sigma_{\rm src}=0.3$, the selected 10 ms protocol
achieved mean held-out one-shot correctness of $0.778$,
compared with $0.609$ at 50 ms. Shortening the total
sensing duration therefore helped at both tested gate
spacings: the benefit was not specific to the original
1~ms spacing.

\subsection{Effective couplings and feasibility}
\label{app:parameter_interpretation}

The normalized waveform makes $g_{\rm sensor}$ an effective
angular-frequency scale per unit waveform amplitude.
The QCS model does not prescribe a sensor platform: a
physical implementation must provide a controllable
signal-sensor interaction with the assumed effective drive
strength and waveform dependence. By contrast,
$g_{\rm coil}$ is an end-to-end coil-output amplitude-transfer
coefficient; its numerical value is not directly comparable
with $g_{\rm sensor}$. Converting $g_{\rm coil}$ to a physical
sensitivity would require the coil geometry, bandwidth,
receiver response, and noise calibration.

For the interaction studied here, the complex molecular
envelope supplies transverse drive quadratures in the
sensor's rotating frame. Representing the molecular signal
in a proton rotating frame does not itself implement that
interaction. The means of obtaining an effective drive, and
its transfer efficiency, phase response, and added noise,
would depend on the sensor platform.

NV centers provide one illustrative example. At the
high-field mixture benchmark's $B_0=11.74~\mathrm{T}$,
the proton carrier is near $500~\mathrm{MHz}$, while the
NV-electron Zeeman frequency scale at the same field is
approximately $330~\mathrm{GHz}$. An NV realization would
therefore require an appropriate signal-to-sensor interface
or high-field sensing protocol; neither is specified by
the present model.

For orientation, define an \emph{illustrative
NV-electron field-equivalent} amplitude
~\cite{Degen2017QuantumSensing} by
\begin{equation}
B_{\rm ref}^{\rm(eq)}
=
\frac{g_{\rm sensor}}{\gamma_{\rm NV}},
\qquad
\gamma_{\rm NV}\simeq
2\pi\times28.0~\mathrm{GHz/T}.
\label{eq:g-to-Bref}
\end{equation}
Thus, $g_{\rm sensor}=300~\mathrm{rad\,s^{-1}}$
gives $B_{\rm ref}^{\rm(eq)}\simeq1.70~\mathrm{nT}$,
whereas $10^4~\mathrm{rad\,s^{-1}}$ gives
$B_{\rm ref}^{\rm(eq)}\simeq56.8~\mathrm{nT}$.
These values express the assumed effective rotation rates
in field units.

The benchmark assumes a coherent, reproducible molecular
transient and is closer in this respect to volume-limited
microscale NMR than to statistical-polarization nanoscale
NV-NMR. Published microscale NV-NMR measurements provide
relevant context
~\cite{Kehayias2017NanostructuredDiamond,
Glenn2018CASR,Smits2019MicrofluidicDiamond}, but do not
directly establish the combined interface, effective drive strength,
control, and readout resources
required by the protocols studied here.

\bibliography{references}

@article{Poggiali2018OptimalControl,
  author  = {Poggiali, F. and Cappellaro, P. and Fabbri, N.},
  title   = {Optimal Control for One-Qubit Quantum Sensing},
  journal = {Physical Review X},
  volume  = {8},
  pages   = {021059},
  year    = {2018},
  doi     = {10.1103/PhysRevX.8.021059}
}

@article{Meinel2021Heterodyne,
  author  = {Meinel, Jonas and Vorobyov, Vadim and Yavkin, Boris and
             Dasari, Durga and Sumiya, Hitoshi and Onoda, Shinobu and
             Isoya, Junichi and Wrachtrup, J{\"o}rg},
  title   = {Heterodyne sensing of microwaves with a quantum sensor},
  journal = {Nature Communications},
  volume  = {12},
  pages   = {2737},
  year    = {2021},
  doi     = {10.1038/s41467-021-22714-y}
}

@article{Kehayias2017NanostructuredDiamond,
author = {Pauli Kehayias and Andrey Jarmola and Nazanin Mosavian and Ilja Fescenko and Francisco M. Benito and Abdelghani Laraoui and Janis Smits and Lykourgos Bougas and Dmitry Budker and Andrej Neumann and Steven R. J. Brueck and Victor M. Acosta},
title = {Solution Nuclear Magnetic Resonance Spectroscopy on a Nanostructured Diamond Chip},
journal = {Nature Communications},
volume = {8},
pages = {188},
year = {2017},
doi = {10.1038/s41467-017-00266-4},
}

@article{Smits2019MicrofluidicDiamond,
author = {Janis Smits and Joshua T. Damron and Pauli Kehayias and Andrew F. McDowell and Nazanin Mosavian and Ilja Fescenko and Nathaniel Ristoff and Abdelghani Laraoui and Andrey Jarmola and Victor M. Acosta},
title = {Two-Dimensional Nuclear Magnetic Resonance Spectroscopy with a Microfluidic Diamond Quantum Sensor},
journal = {Science Advances},
volume = {5},
number = {7},
pages = {eaaw7895},
year = {2019},
doi = {10.1126/sciadv.aaw7895},
}

@article{Kingma2015Adam,
author = {Diederik P. Kingma and Jimmy Ba},
booktitle = {Adam: A Method for Stochastic Optimization},
journal = {International Conference on Learning Representations},
year = {2015},
eprint = {1412.6980},
archivePrefix = {arXiv},
primaryClass = {cs.LG}
}

@article{Loshchilov2017SGDR,
author = {Ilya Loshchilov and Frank Hutter},
booktitle = {{SGDR}: Stochastic Gradient Descent with Warm Restarts},
journal = {International Conference on Learning Representations},
year = {2017},
eprint = {1608.03983},
archivePrefix = {arXiv},
primaryClass = {cs.LG}
}

@article{Degen2017QuantumSensing,
author = {Christian L. Degen and Friedemann Reinhard and Paola Cappellaro},
title = {Quantum Sensing},
journal = {Reviews of Modern Physics},
volume = {89},
pages = {035002},
year = {2017},
doi = {10.1103/RevModPhys.89.035002},
}

@article{Low2016CompositeGates,
author = {Guang Hao Low and Theodore J. Yoder and Isaac L. Chuang},
title = {Methodology of Resonant Equiangular Composite Quantum Gates},
journal = {Physical Review X},
volume = {6},
pages = {041067},
year = {2016},
doi = {10.1103/PhysRevX.6.041067},
}

@article{Low2017QSP,
author = {Guang Hao Low and Isaac L. Chuang},
title = {Optimal Hamiltonian Simulation by Quantum Signal Processing},
journal = {Physical Review Letters},
volume = {118},
pages = {010501},
year = {2017},
doi = {10.1103/PhysRevLett.118.010501},
}

@article{Haah2019QSPDecomposition,
author = {Jeongwan Haah},
title = {Product Decomposition of Periodic Functions in Quantum Signal Processing},
journal = {Quantum},
volume = {3},
pages = {190},
year = {2019},
doi = {10.22331/q-2019-10-07-190},
}

@article{Dong2021QSPPhases,
author = {Yulong Dong and Xiang Meng and K. Birgitta Whaley and Lin Lin},
title = {Efficient Phase-Factor Evaluation in Quantum Signal Processing},
journal = {Physical Review A},
volume = {103},
pages = {042419},
year = {2021},
doi = {10.1103/PhysRevA.103.042419},
}

@article{Joas2025,
  author    = {T. Joas and F. Ferlemann and R. Sailer and P. J. Vetter and J. Zhang and R. S. Said and T. Teraji and S. Onoda and T. Calarco and G. Genov and M. M. M{\"u}ller and F. Jelezko},
  title     = {High-Fidelity Electron Spin Gates for Scaling Diamond Quantum Registers},
  journal   = {Physical Review X},
  volume    = {15},
  number    = {2},
  pages     = {021069},
  year      = {2025},
  doi       = {10.1103/PhysRevX.15.021069}
}

@article{Jauch2025,
  author    = {Isabell Jauch and A. Skljarow and Thomas Strohm and Florian Dolde and Tino Fuchs and F. Jelezko},
  title     = {Enhancing the Ramsey Contrast of an NV-Ensemble in Diamond Using Quantum Optimal Control},
  journal   = {Quantum Science and Technology},
  volume    = {10},
  number    = {4},
  pages     = {045028},
  year      = {2025},
  doi       = {10.1088/2058-9565/adffb2}
}

@misc{hqs_nmr_parameters,
      title={{HQSpectrum}},
      author={{HQS Quantum Simulations GmbH}},
      year={2025},
      url={https://hqspectrum.cloud.quantumsimulations.de},
}

@misc{Fratus2025NMRQuantumSimulation,
author = {Keith R. Fratus and Nicklas Enenkel and Sebastian Zanker and Jan-Michael Reiner and Michael Marthaler and Peter Schmitteckert},
title = {Can a Quantum Computer Simulate Nuclear Magnetic Resonance Spectra Better than a Classical One?},
OPTyear = {2025},
eprint = {2508.06448},
archivePrefix = {arXiv},
primaryClass = {quant-ph},
OPTdoi = {10.48550/arXiv.2508.06448}
}

@misc{Kannan2026QFS,
author = {Ishaan Kannan and Sridhar Prabhu and Saeed A. Khan and Mandar M. Sohoni and Xingrui Song and Saswata Roy and Alen Senanian and Valla Fatemi and Peter L. McMahon and Jordan Cotler},
title = {Exponential Quantum Advantage for Learning Signals with a Single Qubit},
OPTyear = {2026},
eprint = {2608.13521},
archivePrefix = {arXiv},
primaryClass = {quant-ph},
}

@misc{Sen2026QCS,
author = {Purnendu Sen and Mathieu Ouellet and Saeed A. Khan and Wayne Wang and Sridhar Prabhu and Alen Senanian and William P. Banner and William D. Oliver and Peter L. McMahon},
title = {Quantum Sensors That Compute: Quantum Computational Magnetic-Field Sensing Using a Superconducting Qubit},
OPTyear = {2026},
eprint = {2608.17400},
archivePrefix = {arXiv},
primaryClass = {quant-ph},
OPTdoi = {10.48550/arXiv.2608.17400}
}

@misc{Khan2025QCSAdvantage,
  author        = {Saeed A. Khan and Sridhar Prabhu and Logan G. Wright and Peter L. McMahon},
  title         = {Quantum Computational Sensing},
  OPTyear          = {2025},
  eprint        = {2507.16918},
  archivePrefix = {arXiv},
  primaryClass  = {quant-ph},
  OPTdoi           = {10.48550/arXiv.2507.16918}
}

@article{Khan2026QCS,
  author  = {Saeed A. Khan and Sridhar Prabhu and Logan G. Wright and Peter L. McMahon},
  title   = {Quantum Computational Sensing Using Quantum Signal Processing, Quantum Neural Networks, and Hamiltonian Engineering},
  journal = {npj Quantum Information},
  volume  = {12},
  pages   = {119},
  year    = {2026},
  doi     = {10.1038/s41534-026-01235-w}
}

@misc{Prabhu2026QCDS,
  author        = {Sridhar Prabhu and Saeed A. Khan and Xingrui Song and Mathieu Ouellet and Ryotatsu Yanagimoto and Saswata Roy and Alen Senanian and Logan G. Wright and Valla Fatemi and Peter L. McMahon},
  title         = {Quantum Computational Displacement Sensing},
  OPTyear          = {2026},
  eprint        = {2604.13177},
  archivePrefix = {arXiv},
  primaryClass  = {quant-ph}
}

@misc{allen2025quantum_computing_enhanced_sensing,
  author        = {Allen, Richard R. and Machado, Francisco and Chuang, Isaac L. and Huang, Hsin-Yuan and Choi, Soonwon},
  title         = {Quantum Computing Enhanced Sensing},
  OPTyear          = {2025},
  eprint        = {2501.07625},
  archivePrefix = {arXiv},
  primaryClass  = {quant-ph},
  OPTdoi           = {10.48550/arXiv.2501.07625}
}

@article{staudacher2013five_nanometer_nmr,
  author  = {Staudacher, T. and Shi, F. and Pezzagna, S. and Meijer, J. and Du, J. and Meriles, C. A. and Reinhard, F. and Wrachtrup, J.},
  title   = {Nuclear magnetic resonance spectroscopy on a (5-nanometer)$^3$ sample volume},
  journal = {Science},
  year    = {2013},
  volume  = {339},
  number  = {6119},
  pages   = {561--563},
  doi     = {10.1126/science.1231675}
}

@article{Mamin2013,
  author  = {Mamin, H. J. and Kim, M. and Sherwood, M. H. and Rettner, C. T. and Ohno, K. and Awschalom, D. D. and Rugar, D.},
  title   = {Nanoscale Nuclear Magnetic Resonance with a Nitrogen-Vacancy Spin Sensor},
  journal = {Science},
  year    = {2013},
  volume  = {339},
  number  = {6119},
  pages   = {557--560},
  doi     = {10.1126/science.1231540}
}

@article{Schirhagl2014,
  author  = {Schirhagl, Romana and Chang, Kevin and Loretz, Michael and Degen, Christian L.},
  title   = {Nitrogen-Vacancy Centers in Diamond: Nanoscale Sensors for Physics and Biology},
  journal = {Annual Review of Physical Chemistry},
  year    = {2014},
  volume  = {65},
  pages   = {83--105},
  doi     = {10.1146/annurev-physchem-040513-103659}
}

@article{Schmitt2017Qdyne,
  author  = {Schmitt, Simon and Gefen, Tuvia and St{\"u}rner, Felix M. and Unden, Thomas and Wolff, Gerhard and M{\"u}ller, Christoph and Scheuer, Jochen and Naydenov, Boris and Markham, Matthew and Pezzagna, Sebastien and Meijer, Jan and Schwarz, Ilai and Plenio, Martin and Retzker, Alex and McGuinness, Liam P. and Jelezko, Fedor},
  title   = {Submillihertz magnetic spectroscopy performed with a nanoscale quantum sensor},
  journal = {Science},
  year    = {2017},
  volume  = {356},
  number  = {6340},
  pages   = {832},
  doi     = {10.1126/science.aam5532}
}

@article{Glenn2018CASR,
  author  = {Glenn, David R. and Bucher, Dominik B. and Lee, Junghyun and Lukin, Mikhail D. and Park, Hongkun and Walsworth, Ronald L.},
  title   = {High-resolution magnetic resonance spectroscopy using a solid-state spin sensor},
  journal = {Nature},
  year    = {2018},
  volume  = {555},
  pages   = {351--354},
  doi     = {10.1038/nature25781}
}

@article{Aslam2017Memory,
  author  = {Pfender, Matthias and Aslam, Nabeel and Sumiya, Hitoshi and Onoda, Shinobu and Neumann, Philipp and Isoya, Junichi and Meriles, Carlos A. and Wrachtrup, J{\"o}rg},
  title   = {Nonvolatile nuclear spin memory enables sensor-unlimited nanoscale spectroscopy of small spin clusters},
  journal = {Nature Communications},
  year    = {2017},
  volume  = {8},
  pages   = {834},
  doi     = {10.1038/s41467-017-00964-z}
}

@article{zaiser2016enhancing,
  title     = {Enhancing quantum sensing sensitivity by a quantum memory},
  author    = {Zaiser, Sebastian and Rendler, Torsten and Jakobi, Ingmar and Wolf, Thomas and Lee, Sang-Yun and Wagner, Samuel and Bergholm, Ville and Schulte-Herbr{\"u}ggen, Thomas and Neumann, Philipp and Wrachtrup, J{\"o}rg},
  journal   = {Nature Communications},
  volume    = {7},
  pages     = {12279},
  year      = {2016},
  month     = {aug},
  doi       = {10.1038/ncomms12279},
  publisher = {Nature Publishing Group}
}

\end{document}